\documentclass[aps,prx,reprint,superscriptaddress,longbibliography,nofootinbib,amsmath,amssymb,floatfix]{revtex4-2}

\usepackage{graphicx}
\usepackage{bm}
\usepackage{xcolor}
\usepackage{booktabs}
\usepackage[colorlinks=true,linkcolor=blue,citecolor=blue,urlcolor=blue]{hyperref}

\begin{document}
\raggedbottom

\title{Hidden Heat-Current Transport Channels Behind Observable Temperature Dynamics}

\author{Yi Zeng}
\email{yi.zeng@nlr.gov}
\affiliation{National Laboratory of the Rockies, Golden, Colorado 80401, USA}

\author{Jianjun Dong}
\email{dongjia@auburn.edu}
\affiliation{Department of Physics, Auburn University, Auburn, Alabama 36830, USA}

\date{\today}

\begin{abstract}
Diffusive decay and wave-like oscillations in temperature dynamics are commonly used to characterize heat conduction, yet these signatures do not uniquely identify the underlying heat-current transport mechanisms. Here we reveal a spectrum of constitutive heat-current channels whose microscopic organization can change without a corresponding qualitative change in observable temperature dynamics. We develop a two-level spectral theory that first resolves the finite-wave-vector constitutive response into relaxational and streaming channels characterized by their poles and heat-current residues, and then connects their combined response to equilibrium temperature fluctuations through energy conservation. First-principles calculations for NaF at 17~K reveal that the same streaming-induced mode coupling responsible for the perturbative nonlocal correction in the Guyer--Krumhansl (G--K) reduction drives a nonperturbative transformation of the isolated momentum-organized relaxational channel into a dominant complex-conjugate streaming pair. This transformation breaks down the single-relaxational-channel constitutive structure of G--K, while oscillatory temperature dynamics persist below the short-period boundary of its predicted oscillatory window. At shorter periods, heat-current weight redistributes among multiple collision-coupled streaming channels before approaching independently damped phonon streaming. Increasing temperature provides a distinct route beyond hydrodynamics by eliminating the isolated momentum-organized sector, even as oscillatory temperature dynamics persist. Thus, the breakdown of a hydrodynamic constitutive description can reflect a reorganization of heat-current transport rather than the disappearance of oscillatory temperature dynamics, and agreement with temperature measurements alone does not validate the constitutive structure of a reduced theory.
\end{abstract}

\maketitle
\section{Introduction}
\label{sec:introduction}
\begin{figure*}[t!]
\centering
\includegraphics[width=\textwidth]{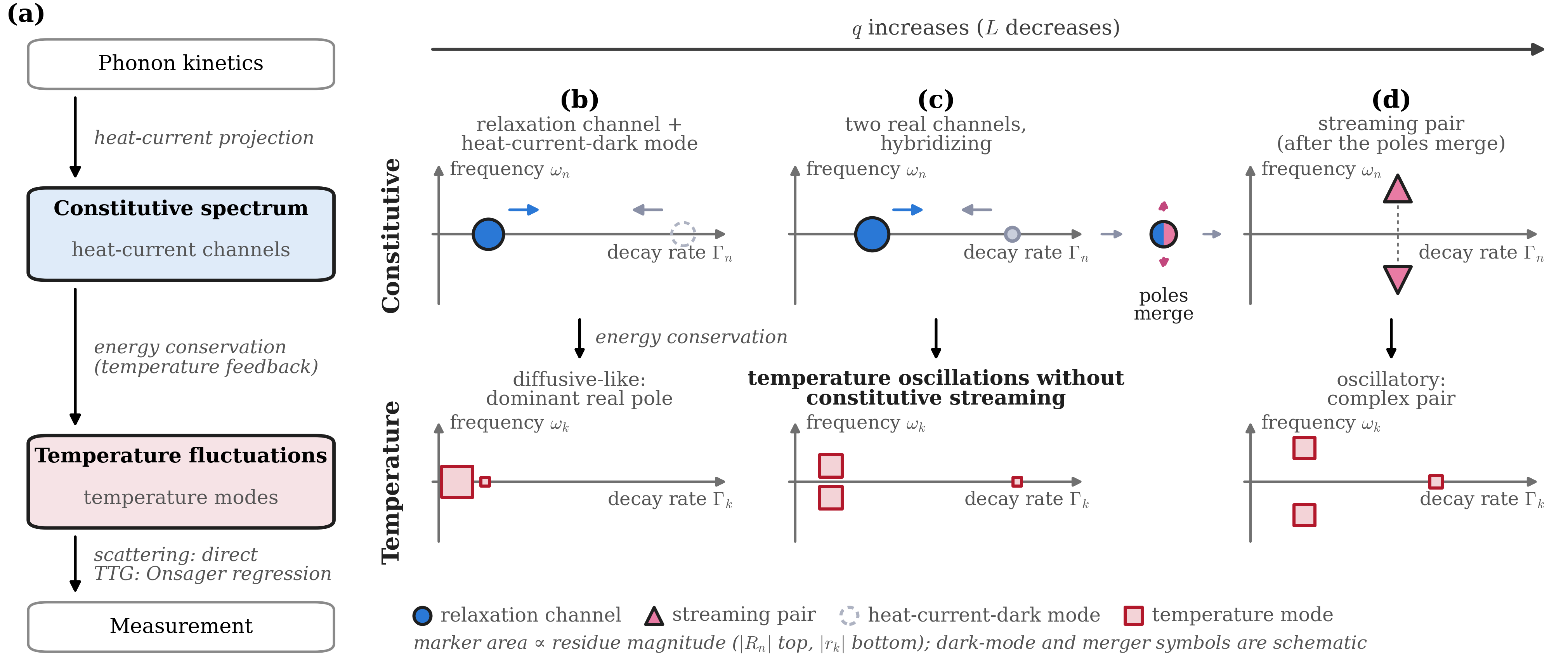}
\caption{\textbf{Constitutive and temperature-fluctuation spectra of non-Fourier heat conduction.} (a) Two-level spectral description connecting microscopic phonon kinetics to observable temperature dynamics through the constitutive heat-current response and energy conservation. (b)--(d) Schematic spectral evolution with increasing wave vector $q$. A dominant relaxational constitutive channel hybridizes with a heat-current-dark kinetic mode and eventually merges with it into a complex-conjugate streaming pair. Temperature modes can nevertheless become oscillatory while the constitutive channels remain relaxational. Horizontal and vertical axes denote positive decay rates and signed frequencies, respectively. Marker areas indicate the magnitudes of constitutive heat-current residues (upper row) or temperature-response residues (lower row).}
\label{fig:schematic}
\end{figure*}

Fourier's law relates heat current instantaneously and locally to the temperature gradient and, together with energy conservation, predicts diffusive temperature evolution~\cite{carslaw1959conduction}. Beyond this description, heat conduction can exhibit temporal memory, spatial nonlocality, collective propagation, and ballistic transport~\cite{cahill2003nanoscale,siemens2010quasi,alvarez2007memory,chen2021non}.
These effects are increasingly relevant to thermal management in highly integrated electronic devices, where shrinking dimensions and localized heating challenge conventional descriptions of heat conduction~\cite{cahill2014nanoscale, pop2010energy}.
Ultrafast and nanoscale thermal measurements have revealed such non-Fourier behavior across a growing range of materials and conditions~\cite{johnson2013direct,johnson2015non,vega2016thermal,2019_EUV_TTG,huberman2019observation,machida2020phonon,PhysRevLett.127.085901,beardo2021observation,ding2022observation,xie2026room,varghese2026controllable}. In particular, transient thermal grating (TTG) experiments probe finite-wave-vector temperature dynamics by monitoring the evolution of a spatially periodic temperature perturbation with period $L$ and wave vector $q=2\pi/L$~\cite{johnson2013direct}. These measurements have revealed length-scale-dependent deviations from Fourier diffusion in dielectric materials~\cite{johnson2013direct,johnson2015non,vega2016thermal,2019_EUV_TTG}, as well as damped oscillatory temperature responses in graphite at temperatures extending to room temperature~\cite{huberman2019observation,ding2022observation,xie2026room}. These observations raise the question of which microscopic transport mechanisms govern temperature evolution beyond Fourier diffusion.

Oscillatory temperature dynamics depart qualitatively from diffusion. Such oscillations are commonly associated with second sound, the wave-like propagation of temperature disturbances. Phonon-mediated second sound was first observed in solid $^4$He and $^3$He~\cite{ackerman1966second,ackerman1969second}, and subsequently in dielectric crystals such as NaF through heat-pulse~\cite{mcnelly1970heat,jackson1970second,jackson1971thermal} and light-scattering measurements~\cite{pohl1976observation}. These observations have historically been interpreted within the framework of phonon hydrodynamics, in which momentum-conserving phonon collisions organize heat transport into collective motion~\cite{peierls1929kinetischen,ward1951velocity,ward1952iii,sussmann1963thermal,chester1963second,guyer1966solution,guyer1966thermal,hardy1970phonon,hardy1971velocity, hardy1974hydrodynamic}. Although temperature oscillations indicate wave-like dynamics, they do not by themselves identify the microscopic heat-current mechanism or establish momentum-organized phonon hydrodynamics.

A broader theoretical challenge is to describe the different regimes of non-Fourier heat conduction within a unified microscopic framework. Existing approaches typically employ reduced descriptions of phonon kinetics, each emphasizing particular transport mechanisms. In the hydrodynamic regime, the Guyer--Krumhansl (G--K) theory exploits the slow relaxation of momentum-organized phonon dynamics to derive a constitutive equation for heat current that incorporates temporal relaxation and spatial nonlocality~\cite{guyer1966solution,guyer1966thermal,hardy1970phonon,hardy1974hydrodynamic}. This reduction retains the dominant momentum-related degree of freedom while incorporating the effects of faster kinetic modes through effective nonlocal transport coefficients. Generalized G--K-type descriptions introduce modified memory and nonlocal transport coefficients to extend their applicability beyond conventional phonon hydrodynamics~\cite{alvarez2018thermal,sendra2021derivation,beardo2021general,guo2018phonon,tur2026high}. Related formulations, including the viscous heat equations, describe collective and nonlocal transport through coupled temperature and phonon drift-velocity fields~\cite{simoncelli2020generalization}. In contrast to the momentum-organized description underlying G--K, the relaxation-time approximation (RTA) treats heat-current transport as distributed among independently relaxing phonon populations streaming at their group velocities, replacing the coupled collision operator with mode-dependent relaxation rates~\cite{born1954dynamical,cowley1963lattice,maradudin1962scattering,Tang4539,tang2014thermal}. It provides a widely used description of ballistic, quasiballistic, and nonlocal heat transport~\cite{mahan1988nonlocal,minnich2012determining,koh2014nonlocal,regner2013broadband,vermeersch2015superdiffusive,allen2018analysis,hua_minnich_2018_PRB,hua2019generalized}. These approaches thus represent contrasting organizations of microscopic heat-current transport, from a dominant momentum-organized relaxation channel to distributed phonon streaming, raising the question of how transport evolves between them with temperature and wave vector, and when their assumed constitutive structures remain valid.

First-principles calculations based on the phonon Boltzmann transport equation (BTE) retain the microscopic collision and streaming dynamics without imposing a reduced constitutive law. Such calculations have been used to determine thermal conductivity~\cite{togo_phono3py}, parameters of reduced transport models~\cite{simoncelli2020generalization,albert2026phonon}, and non-Fourier temperature responses~\cite{lee2015hydrodynamic,cepellotti2015phonon,li2019crossover,chiloyan2021green,kremeyer2026transition}. Complementary developments have characterized the eigenmodes of the full finite-wave-vector kinetic generator as transport waves~\cite{cepellotti2017transport} and derived the space-time-dependent thermal conductivity as an aggregate constitutive response~\cite{hua2020space}. While these approaches retain the underlying kinetic information, the full kinetic spectrum does not directly isolate the channels of the constitutive heat-current response, and the aggregate conductivity does not explicitly reveal their individual contributions. Identifying the microscopic organization of heat-current transport requires resolving the finite-wave-vector constitutive spectrum through both its dynamical poles and heat-current residues.

Here we develop a two-level spectral theory by resolving the causal heat-current kernel of our earlier work~\cite{crawford2024theory,zeng_and_dong_2026_unified} into finite-wave-vector transport channels. The poles specify the channels' relaxational and streaming dynamics, and the residues specify their contributions to heat current. Energy conservation couples their combined constitutive response to temperature, producing a distinct spectrum of temperature poles and residues. This spectrum governs equilibrium temperature fluctuations and, through Onsager regression~\cite{onsager1931reciprocal}, the relaxation of small, locally equilibrated temperature modulations. Figure~\ref{fig:schematic} illustrates the distinction: relaxational constitutive channels can produce oscillatory temperature dynamics, whereas streaming-dominated heat-current transport need not produce temperature sign reversal.

First-principles calculations for NaF reveal two distinct routes beyond hydrodynamic heat transport. At 17~K, where the momentum-organized sector remains well isolated at $q=0$, increasing wave vector drives a fundamental reorganization of the constitutive spectrum. The same streaming-induced coupling responsible for the perturbative nonlocal correction in the G--K reduction transforms the isolated momentum-organized relaxational channel, through hybridization with an initially heat-current-dark kinetic mode, into a dominant complex-conjugate streaming pair. Although this transformation breaks down the single-relaxational-channel constitutive structure of G--K, the transformed pair continues to dominate heat-current transport, and oscillatory temperature dynamics persist below the short-period boundary of the G--K oscillatory window. At shorter periods, heat-current weight redistributes among multiple collision-coupled streaming channels before approaching independently damped phonon streaming. Increasing temperature provides a distinct route beyond hydrodynamics by eliminating the isolated momentum-organized sector itself, while oscillatory temperature dynamics can persist through other collision-coupled channels.

\section{Theory}
\label{sec:theory}

\subsection{Constitutive spectrum}
\label{sec:constitutive_spectrum}

We begin with the causal constitutive relation describing the heat-current response to a prescribed temperature-gradient field~\cite{zeng_and_dong_2026_unified},
\begin{equation}
\mathbf j(\mathbf r,t)=-\int d\mathbf r'\int_{-\infty}^{t}dt'\,\overleftrightarrow{Z}(\mathbf r,\mathbf r',t-t')\cdot\boldsymbol{\nabla}T(\mathbf r',t'),
\label{eq:kernel}
\end{equation}
where $\overleftrightarrow{Z}$ generalizes thermal conductivity to spatially nonlocal and history-dependent transport. The kernel characterizes heat-current response to thermal driving before the feedback imposed by energy conservation is included. For a homogeneous medium, $\overleftrightarrow{Z}(\mathbf r,\mathbf r',t)=\overleftrightarrow{Z}(\mathbf r-\mathbf r',t)$, and its Fourier--Laplace transform is
\begin{equation}
\overleftrightarrow{\tilde Z}(\mathbf q,s)=\int_0^\infty dt\,e^{-st}\int d\mathbf r\,e^{-i\mathbf q\cdot\mathbf r}\overleftrightarrow{Z}(\mathbf r,t).
\label{eq:kernel_transform}
\end{equation}
The constitutive relation becomes $\tilde{\mathbf j}(\mathbf q,s)=-\overleftrightarrow{\tilde Z}(\mathbf q,s)\cdot i\mathbf q\,\delta\tilde T(\mathbf q,s)$. Temperature evolution couples to the heat-current component along $\mathbf q$. Therefore, we define the longitudinal current $\tilde j_L=\hat{\mathbf q}\cdot\tilde{\mathbf j}$ and longitudinal kernel $\tilde Z_L=\hat{\mathbf q}\cdot\overleftrightarrow{\tilde Z}\cdot\hat{\mathbf q}$, giving $\tilde j_L=-iq\tilde Z_L\,\delta\tilde T$, where $q=|\mathbf q|$. The corresponding real-frequency kernel is a generalized nonlocal, frequency-dependent thermal conductivity~\cite{hua2020space}.

For dielectric crystals, the constitutive kernel follows from the linearized phonon BTE. In the symmetrized phonon basis, the kinetic generator is $\boldsymbol{\mathcal K}(\mathbf q)=\boldsymbol\Omega+i\mathbf q\cdot\mathbf V$, where $\boldsymbol\Omega$ is the symmetrized collision operator and $\mathbf V$ is the diagonal operator of phonon group velocities~\cite{zeng_and_dong_2026_unified}~(Appendix~\ref{app:methods_projected_transport}). Energy conservation defines a normalized null vector $|e\rangle$ satisfying $\boldsymbol\Omega|e\rangle=0$. The orthogonal projector $\mathbf Q=\mathbf I-|e\rangle\langle e|$ separates this conserved energy mode from the remaining kinetic degrees of freedom. Its range, $\operatorname{Ran}\mathbf Q$, is the subspace orthogonal to $|e\rangle$. The \emph{transport generator} is the restriction of the projected kinetic operator to this subspace:
\begin{equation}
\boldsymbol{\Xi}(\mathbf q) = \left. \mathbf Q(\boldsymbol\Omega+i\mathbf q\cdot\mathbf V)\mathbf Q \right|_{\operatorname{Ran}\mathbf Q}. \label{eq:transport_generator}
\end{equation}
The generator is complex symmetric, $\boldsymbol\Xi^{\mathsf T}=\boldsymbol\Xi$, but generally non-Hermitian. For an inversion-symmetric crystal, the parity operator $\boldsymbol{\mathcal P}$ satisfies $\boldsymbol{\mathcal P}\boldsymbol\Xi(\mathbf q)\boldsymbol{\mathcal P}=\boldsymbol\Xi(\mathbf q)^*$, ensuring that its eigenvalues are either real or occur in complex-conjugate pairs. The eigenvalues $a_n=\Gamma_n+i\omega_n$ specify the relaxation rates $\Gamma_n$ and oscillation frequencies $\omega_n$ of the projected kinetic modes. The positive semidefiniteness of the collision operator guarantees $\operatorname{Re}a_n\geq0$. The right and left eigenvectors of $\boldsymbol\Xi$ satisfy $\boldsymbol\Xi|\phi_n^{\mathrm R}\rangle=a_n|\phi_n^{\mathrm R}\rangle$ and $\langle\phi_n^{\mathrm L}|\boldsymbol\Xi=a_n\langle\phi_n^{\mathrm L}|$, respectively, with biorthogonal normalization $\langle\phi_m^{\mathrm L}|\phi_n^{\mathrm R}\rangle=\delta_{mn}$. 

We define the heat-current coupling vector $|\Lambda\rangle=\mathbf V|e\rangle$, orthogonal to $|e\rangle$ by inversion symmetry, and the volumetric heat capacity $c_V=C/V$, where $C$ is the total heat capacity and $V$ is the crystal volume. The constitutive kernel has the spectral representation (Appendix~\ref{app:kernel}):
\begin{equation}
\tilde Z_{ij}(\mathbf q,s)=c_V\sum_n\frac{\langle\Lambda_i|\phi_n^{\mathrm R}\rangle\langle\phi_n^{\mathrm L}|\Lambda_j\rangle}{s+a_n(\mathbf q)}=\sum_n\frac{R_{ij}^{(n)}(\mathbf q)}{s+a_n(\mathbf q)}.
\label{eq:constitutive_spectrum_general}
\end{equation}
For the longitudinal component, we write $R_n(q)\equiv R_L^{(n)}(q)=\hat q_iR_{ij}^{(n)}(\mathbf q)\hat q_j$. Each eigenmode with a real eigenvalue, or pair of eigenmodes with complex-conjugate eigenvalues, defines a \emph{transport channel} when its heat-current residue is nonzero. Conjugate partners are counted separately in the spectral sums.

A real eigenvalue $a_n=\Gamma_n$ defines a \emph{relaxational channel}, whose time-domain contribution is
\begin{equation}
z_n^{R}(t)=R_n e^{-\Gamma_n t}.
\label{eq:relaxational_channel_time}
\end{equation}
A complex-conjugate pair $a_{n,\pm}=\Gamma_n\pm i\omega_n$ defines a \emph{streaming channel}, whose combined contribution is
\begin{equation}
\begin{aligned}
z_n^{S}(t)&=2e^{-\Gamma_n t}\operatorname{Re} \left[R_n e^{-i\omega_n t}\right]\\
&=2|R_n|e^{-\Gamma_n t}\cos(\omega_n t-\arg R_n),
\end{aligned}
\label{eq:streaming_channel_time}
\end{equation}
where the constitutive poles are at $s=-a_n$, and $R_n$ is the residue of one member of the conjugate pair. The eigenvalues $a_n$, together with their heat-current residues $R_{ij}^{(n)}$, specify the \emph{constitutive spectrum}.

Transport channels can have signed or complex heat-current residues, so their contributions can reinforce or cancel. Their total spectral weight nevertheless obeys the wave-vector-independent sum rule
\begin{equation}
\sum_n R_{ij}^{(n)}(\mathbf q)=Z_{ij}(\mathbf q,0^+)=c_V\langle e| \mathbf V_i \mathbf V_j|e\rangle.
\label{eq:constitutive_sum_rule}
\end{equation}
The sum rule becomes $\sum_n R_L^{(n)}(q)=c_Vv_{\rm rms}^2$, where $v_{\rm rms}^2=\langle e|\mathbf V_L^2|e\rangle$ and $\mathbf V_L=\hat{\mathbf q}\cdot\mathbf V$. Here $v_{\rm rms}$ is the heat-capacity-weighted root-mean-square velocity component along $\mathbf q$.

In contrast, the static constitutive response is
\begin{equation}
\kappa_{ij}(\mathbf q)=\tilde Z_{ij}(\mathbf q,0) =\sum_n\frac{R_{ij}^{(n)}(\mathbf q)}{a_n(\mathbf q)}.
\label{eq:static_constitutive_spectrum}
\end{equation}
Whereas the instantaneous response is constrained by the total spectral weight, the static response weights each residue by $1/a_n$. Consequently, the static conductivity alone cannot determine the underlying distribution of relaxation rates, streaming frequencies, and heat-current residues.

To visualize the constitutive spectrum, we introduce the normalized real-frequency representation
\begin{equation}
S_Z(q,u)=\frac{2q}{\pi c_Vv_{\rm rms}^2}\operatorname{Re}\tilde Z_L(q,iqu),
\label{eq:constitutive_speed_spectrum}
\end{equation}
where $u=\omega/q$ is the phase-speed variable. Using $u=\omega/q$ lets us compare spectra across wave vectors and with phonon group velocities, particularly when streaming frequencies scale approximately linearly with $q$. For an inversion-symmetric crystal, $S_Z(q,u)$ is nonnegative and satisfies $\int_0^\infty S_Z(q,u)\,du=1$ (Appendix~\ref{app:spectral}). This continuous representation sums the constitutive poles and residues into one profile, so a maximum at finite $u$ can come from several overlapping channels as well as from a single propagating eigenmode. At fixed temperature, increasing $q$ strengthens the streaming term $iq\mathbf V_L$ relative to the collision operator. The NaF calculations determine how this competition redistributes heat-current weight among transport channels.

\subsection{Equilibrium temperature-fluctuation spectrum}
\label{sec:fluctuations}

Energy conservation couples the transport channels to the conserved energy coordinate, producing a second spectral level associated with equilibrium temperature fluctuations. Writing $\delta T=\delta e/c_V$, with $\delta e$ the energy-density fluctuation, and defining $\delta T(\mathbf q)=V^{-1}\int d\mathbf r\,e^{-i\mathbf q\cdot\mathbf r}\delta T(\mathbf r)$, we introduce the volume-normalized equilibrium autocorrelation
\begin{equation}
K_{TT}(q,t)\equiv V\langle\delta T(\mathbf q,t)\delta T(-\mathbf q,0)\rangle_{\rm eq}=\frac{k_{\rm B}T_0^2}{c_V}F_T(q,t),
\label{eq:KTT_main}
\end{equation}
where $T_0$ is the equilibrium temperature and $F_T(q,0)=1$. The normalization gives $K_{TT}$ a finite thermodynamic limit.

Within linearized phonon kinetics, the normalized correlation $F_T$ is the energy-sector matrix element of the full kinetic propagator. Eliminating the energy-orthogonal subspace through the Schur complement of $s\mathbf I+\boldsymbol{\mathcal K}$ gives the memory-function form (Appendix~\ref{app:methods_temperature_response})
\begin{equation}
\begin{aligned}
\tilde F_T(q,s)
&=\langle e|[s\mathbf I+\boldsymbol{\mathcal K}(\mathbf q)]^{-1}|e\rangle\\
&=\frac{1}{s+q^2\tilde Z_L(q,s)/c_V}\\
&=\frac{1}{s+q^2\sum_nR_n(q)/\{c_V[s+a_n(q)]\}}.
\end{aligned}
\label{eq:temperature_green_main}
\end{equation}
Resolving this standard memory-function relation~\cite{kadanoff1963hydrodynamic,zwanzig1961memory,mori1965transport} into pole--residue channels connects the heat-current dynamics generated by $\boldsymbol\Xi$ to the temperature modes of $\boldsymbol{\mathcal K}$.

The temperature-fluctuation spectrum is obtained from the poles and residues of the full energy-sector resolvent,
\begin{equation}
\tilde F_T(q,s)=\sum_k\frac{r_k(q)}{s-p_k(q)}, \qquad \sum_k r_k(q)=1, \label{eq:temperature_pole_spectrum}
\end{equation}
where $p_k(q)$ are the temperature poles and $r_k(q)$ their normalized residues. Only full kinetic eigenmodes with nonzero energy-sector residues contribute to this spectrum. The poles generally differ from the constitutive poles $-a_n(q)$ and are determined by the zeros of the secular denominator in Eq.~\eqref{eq:temperature_green_main}. Their negatives, $-p_k$, are the eigenvalues of these modes, which form the energy-carrying subset of the transport waves of Cepellotti and Marzari~\cite{cepellotti2017transport}.

Through Onsager regression~\cite{onsager1931reciprocal,mori1965transport,zwanzig1961memory}, the normalized equilibrium autocorrelation also describes the relaxation of a small, locally equilibrated temperature modulation,
\begin{equation}
G_T(q,t)=\frac{K_{TT}(q,t)}{K_{TT}(q,0)}=F_T(q,t).
\label{eq:Onsager_regression_main}
\end{equation}
Transient thermal gratings probe this relaxation, while equilibrium scattering measurements can access the corresponding fluctuation spectrum through an appropriate optical coupling. With our Fourier convention, the dynamic temperature-fluctuation spectrum is $S_{TT}(q,\omega)=2K_{TT}(q,0)\operatorname{Re}\tilde F_T(q,i\omega)$. These measurements probe the temperature-fluctuation spectrum, whose poles and residues are generated by the full kinetic evolution.

In parallel with $S_Z(q,u)$, we define the normalized real-frequency representation
\begin{equation}
S_T(q,u)=\frac{2q}{\pi}\operatorname{Re}\tilde F_T(q,iqu).
\label{eq:temperature_speed_spectrum}
\end{equation}
For an inversion-symmetric crystal, $S_T(q,u)$ is likewise nonnegative and satisfies $\int_0^\infty S_T(q,u)\,du=1$. The two representations place constitutive heat-current response and equilibrium temperature fluctuations on the same velocity scale, while retaining their distinct pole-residue structures. Because $\tilde F_T$ depends on the kernel only through the term $q^2\tilde Z_L/c_V$ added to $s$, $F_T$ is insensitive to the kernel at frequencies where that term is small compared with $|s|$, and at long times it depends mainly on the low-frequency kernel. A reduced model can closely approximate a temperature trace with an inaccurate kernel.

\subsection{Reduced transport descriptions}
\label{sec:limiting}

Fourier theory, the G--K reduction, RTA, and the collisionless limit provide reference descriptions of the two spectral levels. They differ in the constitutive dynamics retained and in the resulting organization of heat-current transport and temperature fluctuations.

\emph{Fourier diffusion.} Let $\kappa=\tilde Z_L(0,0)$ and $D_0=\kappa/c_V$ denote the bulk conductivity and diffusivity. Fourier's law is recovered in the long-time, long-wavelength limit, when the heat-current-carrying channels that contribute appreciably respond on timescales short compared with temperature evolution. The constitutive response is then approximately independent of frequency and approaches the static nonlocal conductivity, $\tilde Z_L(q,s)\simeq\kappa(q)$. When spatial nonlocality is also negligible, $\kappa(q)\simeq\kappa$, giving $\tilde Z_L^{\rm F}(q,s)=\kappa$ or $Z_L^{\rm F}(q,t)=\kappa\,\delta(t)$. This limit applies to both relaxational and streaming channels: their microscopic dynamics are unresolved on the diffusive timescale $(D_0q^2)^{-1}$. The microscopic kernel has finite initial weight, whereas the Fourier kernel is singular at $t=0$ and has no normalized constitutive spectrum of the form in Eq.~\eqref{eq:constitutive_speed_spectrum}.

Energy conservation then gives $\tilde F_T^{\rm F}(q,s)=(s+D_0q^2)^{-1}$. The corresponding normalized temperature-fluctuation spectrum is a Lorentzian,
\begin{equation}
S_T^{\rm F}(q,u)=\frac{2}{\pi}\frac{D_0q}{u^2+(D_0q)^2}.
\label{eq:Fourier_spectrum}
\end{equation}
Fourier diffusion has no resolved constitutive memory but a finite-rate temperature relaxation.

\emph{Guyer--Krumhansl reduction.} The G--K kernel~\cite{guyer1966solution,guyer1966thermal,sendra2022hydrodynamic} is
\begin{equation}
\tilde Z_L^{\rm GK}(q,s)=\frac{\kappa}{1+\tau s+\ell^2q^2}=\frac{\kappa\Gamma_h}{s+\gamma_q},
\label{eq:GK_kernel_main}
\end{equation}
where $\Gamma_h=\tau^{-1}$, $\mu=\ell^2/\tau$, and $\gamma_q=\Gamma_h+\mu q^2$. The G--K reference retains a single momentum-organized relaxational channel with pole $s=-\gamma_q$ and residue $\kappa\Gamma_h$. This conductivity normalization reproduces the exact bulk diffusivity $D_0$, but its residue differs from the microscopic hydrodynamic-channel value $\kappa_h\Gamma_h$ unless $\kappa_h = \kappa$, where $\kappa_h$ is that channel's contribution to the bulk conductivity. The parameters $\Gamma_h$ and $\mu$ follow from the isolated slow sector of the $q=0$ collision spectrum and its second-order streaming correction. The reduction is controlled while the momentum-organized sector remains spectrally isolated and streaming-induced coupling to other kinetic modes is perturbatively weak.

Coupling the constitutive response to energy conservation yields the temperature-fluctuation propagator
\begin{equation}
\tilde F_T^{\rm GK}(q,s)=\frac{s+\gamma_q}{s^2+\gamma_qs+q^2v_h^2},
\label{eq:GK_FT_main}
\end{equation}
where $v_h^2=D_0\Gamma_h$. Within this reduction, the temperature response becomes oscillatory when the poles form a complex-conjugate pair, requiring
\begin{equation}
\gamma_q<2qv_h.
\label{eq:GK_second_sound_condition}
\end{equation}
The periods satisfying this condition define the G--K oscillatory window, with edges given by Eq.~\eqref{eq:GK_q_bounds}. The corresponding constitutive and temperature-fluctuation spectra are given in Appendix~\ref{app:GK}.

\emph{Relaxation-time approximation.} In the energy-conserving RTA, phonons stream at their group velocities and relax at mode-dependent rates toward a local-equilibrium temperature~\cite{maznev2011onset,hua2014transport,hua2014analytical}. Its collision operator combines diagonal phonon relaxation rates with a rank-one correction that enforces energy conservation (Appendix~\ref{sec:methods_ecRTA}). Unlike the single-channel G--K reduction, RTA retains the distribution of phonon velocities and relaxation times while neglecting the off-diagonal collision couplings responsible for collective transport. The conserving correction nevertheless couples phonon populations through the energy coordinate, so the projected constitutive channels are not generally independent phonon modes. Both spectral responses have analytical expressions in phonon-mode sums. Agreement with RTA indicates that the neglected off-diagonal collision couplings have little influence on the aggregate response.

\emph{Collisionless streaming.} In the collisionless limit, the kinetic generator reduces to $\boldsymbol{\mathcal K}^{\rm cl}(q)=iq\mathbf V_L$, and each phonon mode streams independently at frequency $qv_{\alpha L}$. The normalized temperature correlation and its spectrum are
\begin{equation}
F_T^{\rm cl}(q,t) =\frac{1}{C}\sum_\alpha c_\alpha(T) \cos(qv_{\alpha L}t), \label{eq:collisionless_FT_main}
\end{equation}
\begin{equation}
S_T^{\rm cl}(q,u) =\frac{1}{C}\sum_\alpha c_\alpha(T) \delta(u-|v_{\alpha L}|), \label{eq:collisionless_ST_main}
\end{equation}
where $c_\alpha(T)$ is the mode heat capacity and $C=\sum_\alpha c_\alpha(T)$. The temperature correlation is a superposition of independent streaming contributions that can decay through velocity dephasing without collisions, with or without temperature sign reversal, depending on the heat-capacity-weighted velocity distribution.

The collisionless constitutive kernel is governed by the energy-projected streaming operator, $\boldsymbol\Xi^{\rm cl}(q)=iq\mathbf Q\mathbf V_L\mathbf Q$. The exact relation between the two spectral levels remains valid and gives
\begin{equation}
\begin{aligned}
\tilde Z_L^{\rm cl}(q,s)&=\frac{c_V}{q^2}\left[\frac{1}{\tilde F_T^{\rm cl}(q,s)}-s\right],\\
\tilde F_T^{\rm cl}(q,s)&=\sum_\alpha\frac{c_\alpha(T)}{C}\frac{s}{s^2+q^2v_{\alpha L}^2}.
\end{aligned}
\label{eq:collisionless_Z_time}
\end{equation}
Thus, the constitutive spectrum is determined by the projected operator $\mathbf Q\mathbf V_L\mathbf Q$, rather than directly by the velocity distribution governing temperature fluctuations. The two levels also weight phonons differently: temperature fluctuations carry heat-capacity weights $c_\alpha$, whereas the instantaneous constitutive response carries weights $c_\alpha v_{\alpha L}^2$. Consequently, slow phonons can contribute substantially to temperature fluctuations while contributing little to heat current. For a single pair of velocities $\pm v$, the time-domain constitutive kernel is constant, $Z_L^{\rm cl}(q,t)=c_Vv^2$, or $\tilde Z_L^{\rm cl}(q,s)=c_Vv^2/s$, corresponding to a zero-rate relaxational channel, while energy conservation gives $F_T^{\rm cl}=\cos(qvt)$.

For a fixed harmonic phonon dispersion, the streaming frequencies $qv_{\alpha L}$ are temperature-independent, but the heat-capacity weights vary with temperature. Their redistribution toward slower phonons can suppress temperature sign reversal even in the absence of scattering. The collisionless correlation $F_T^{\rm cl}(q,t)$ depends on $q$ and $t$ only through $qt$, so at fixed temperature it has the same shape in the scaled time $qt$. The collisionless limit thus provides a reference for separating velocity dephasing from scattering and collective collision coupling.

\emph{Reduced-model accuracy.} We denote the validity boundaries of Fourier, G--K, RTA, and collisionless transport by $L_F$, $L_{\rm GK}$, $L_{\rm RTA}$, and $L_{\rm cl}$, respectively, each defined by a 1\% deviation of the reduced constitutive kernel from the full kernel. Appendix~\ref{app:limiting_reductions} derives estimates from the collision operator and phonon velocities, and Sec.~S3\,A specifies the comparison measures and evaluates the estimates against the computed boundaries.

\subsection{Spectral diagnostics}
\label{sec:diagnostics}

We characterize general phonon transport through the pole-residue organization of the constitutive and temperature spectra with numerical thresholds in Sec.~S3.

\emph{Constitutive organization.} A real constitutive pole is one channel with weight $w_n=|R_n|$, and a conjugate pair one channel with $w_n=|R_{n,+}+R_{n,-}|$. Poles that nearly coalesce are combined into one channel of weight $|\sum R_n|$, because their individual residues grow large with opposite signs while their sum stays finite (Appendix~\ref{app:spectral}). Whether the heat-current weight is carried by one channel or distributed over many is measured by the largest grouped share $\hat w_1=\max_nw_n/\sum_mw_m$, with one dominant channel for $\hat w_1\ge1/2$. Whether it relaxes or oscillates is measured by the streaming index
\begin{equation}
W_s=\frac{\sum_nw_n\sigma_n}{\sum_nw_n},
\label{eq:Ws_appendix}
\end{equation}
where $\sigma_n=\omega_n^2/(\Gamma_n^2+\omega_n^2)$ vanishes for a purely relaxational pole and approaches unity for weakly damped oscillation, and $\sigma_n$ of a combined channel is the $|R|$-weighted mean over its members. Relaxation dominates for $W_s<1/2$, and the streaming crossover $L_s$ is the first crossing of $W_s=1/2$ with decreasing period. The diagnostics use the grouped-weight normalization $\sum_m w_m$, while the plotted residues use the signed-sum normalization $\sum_n R_n$.

\emph{Temperature-mode organization.} The temperature poles $p_k=-\Gamma_k\pm i\omega_k$ and residues $r_k$, with $\sum_kr_k=F_T(q,0)=1$, are grouped into channels in the same way. The grouped temperature-channel weights $w_k$ are obtained from the residues $r_k$ using the same rule as for $w_n$. The leading-pair share $P_{\rm pair}$ is the largest pair weight over $\sum_kw_k$. The decay amplitude $P_D$ is the residue of the slowest channel, the one with the smallest $\Gamma_k$, and its rate $\Gamma_D$ defines $D(q)=\Gamma_D/q^2$. For the oscillating channels ($\omega_k>0$),
\begin{equation}
\begin{aligned}
\bar\omega&=\frac{\sum_kw_k\omega_k}{\sum_kw_k},\qquad
\bar\Gamma=\frac{\sum_kw_k\Gamma_k}{\sum_kw_k},\\
\Delta\omega&=\left[\frac{\sum_kw_k(\omega_k-\bar\omega)^2}{\sum_kw_k}\right]^{1/2}.
\end{aligned}
\label{eq:dephasing_ratio}
\end{equation}
The ratio $\Delta\omega/\bar\Gamma$ provides an approximate comparison of the dephasing and damping timescales.

\emph{Temperature-fluctuation classes.} Equilibrium temperature fluctuations are classified through $F_T(q,t)$ and its temperature poles. A response reverses sign when $\min_tF_T(q,t)<0$ over $0\le tqv_{\rm rms}/(2\pi)\le4$. For the temperature map, we call a sign-reversing response dominated by one pole pair ($P_{\rm pair}\ge1/2$) second sound. The remaining sign-reversing responses are classified as dephasing ($\Delta\omega\ge\bar\Gamma$) or damped oscillatory responses ($\Delta\omega<\bar\Gamma$). Without temperature sign reversal, a response is diffusive when its slowest pole carries the full amplitude and decays at the Fourier rate, $|P_D-1|\le1\%$ and $|D(q)/D_0-1|\le1\%$; diffusive-like when $P_D\ge1/2$ but the diffusive criteria are not both met, subdivided by whether $D(q)$ lies above or below $D_0$; and non-diffusive when $P_D<1/2$. Responses with $L<L_{\rm cl}$, where the constitutive kernel agrees with the collisionless reference within 1\%, are ballistic (Appendix~\ref{sec:methods_collisionless}). A collective thermal wave is identified separately, by the weight, damping and frequency concentration of the leading pair and its continuous dispersion across wave vectors (Sec.~S2\,C), so it can appear where the map shows another class.

\section{First-principles results for NaF}
\label{sec:NaF_results}
Here, we use NaF, a benchmark material for phonon hydrodynamics, to study how momentum-organized collective transport evolves into other non-Fourier regimes. Using first-principles phonon kinetics, we track the transport channels and equilibrium temperature fluctuations across temperatures and length scales, from low-temperature hydrodynamics to nanoscale, streaming-dominated transport. 

The harmonic and third-order interatomic force constants for rock-salt NaF were obtained from the NIMS Materials Data Repository~\cite{togo_NaF_MDR}. These force constants were generated using polynomial machine-learning potentials trained on density-functional theory calculations employing the PBEsol exchange-correlation functional~\cite{togo_seko_2024_mlp}. Phonon properties and three-phonon scattering matrix elements were calculated using \texttt{phono3py}~\cite{togo_phono3py} on a $\Gamma$-centered $17\times17\times17$ Brillouin-zone mesh, with the supplied dipole correction included in the harmonic dynamical matrix. Energy-conservation delta functions were evaluated with the linear tetrahedron method, and the three zero-frequency acoustic modes at the zone center were excluded. The calculations describe intrinsic bulk NaF with three-phonon scattering, excluding boundary effects and thermoelastic coupling. Natural NaF is monoisotopic, so isotope-disorder scattering is absent. The characteristic crossover lengths reported below refer to intrinsic bulk NaF and may shift in finite experimental samples due to extrinsic scattering and boundary effects.

The collision operator was constructed as an energy-conserving, positive-semidefinite sum of three-phonon scattering events in Chaput's symmetrized representation~\cite{PhysRevLett.110.265506}. Crystal symmetry was used to reduce the longitudinal kinetic problem for $\mathbf q\parallel[100]$. Diagonalization of the projected generator $\boldsymbol\Xi(q)$ yielded the constitutive poles and residues, while the temperature poles and residues were obtained from the full kinetic generator using the arrowhead representation [Eq.~\eqref{eq:arrowhead_appendix}]. Time-domain correlations were reconstructed from these poles and residues. The G--K, RTA, and collisionless references were constructed from the same phonon inputs, and crossover periods were determined using the spectral diagnostics of Sec.~\ref{sec:diagnostics}. Numerical checks, mesh comparisons, and benchmarks against measured thermal conductivity and second-sound propagation are presented in Sec.~S1 of the Supplemental Material~\cite{SM}, whose sections, figures and tables are labeled with the prefix S.

\subsection{Limits of the hydrodynamic reduction}
\label{sec:hydrodynamic_breakdown}

\begin{figure*}[!t]
\centering
\includegraphics[width=\textwidth]{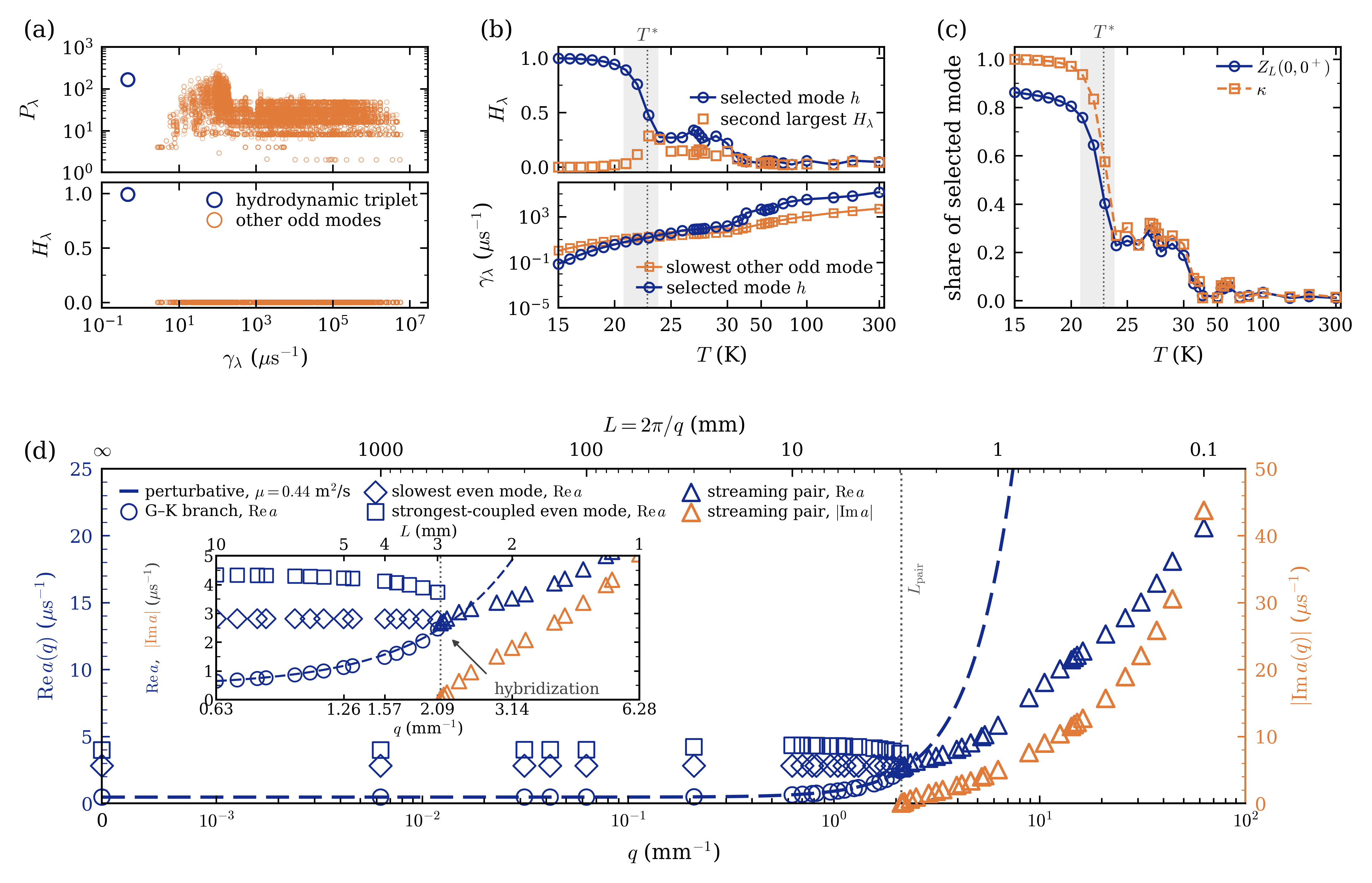}
\caption{\textbf{Momentum-organized hydrodynamics and its temperature- and wave-vector-dependent evolution in NaF.} (a) Participation ratios $P_\lambda$ and momentum overlaps $H_\lambda$ of the odd-parity $q=0$ collision eigenmodes at 17~K. (b) Temperature dependence of the momentum overlap and relaxation rate of the most momentum-aligned mode, selected at each temperature, compared with the second-largest momentum overlap (upper) and the slowest remaining odd-mode relaxation rate (lower). (c) Contributions of this selected mode to the instantaneous constitutive weight $Z_L(0,0^+)$ and bulk thermal conductivity $\kappa$. The temperature axis in (b,c) is linear up to 30~K and logarithmic above. Shading in (b,c) highlights the rapid reorganization of the momentum-organized collision sector, and the dotted line marks $T^*=22.9$~K. Above $T^*$, the selected mode can change between neighboring mixed modes. (d) Finite-wave-vector evolution of the longitudinal constitutive eigenvalues at 17~K, compared with the perturbative G--K dispersion (dashed curve). Circles: hydrodynamic branch; diamonds: the slowest even mode, which carries no heat current at $q=0$ and merges with the hydrodynamic branch; squares: the even mode most strongly coupled to the hydrodynamic branch; triangles: the complex-conjugate streaming pair formed at $L_{\rm pair}$ (dotted line). Relaxation rates $\operatorname{Re}a$ are shown on the left axis and oscillation frequencies $|\operatorname{Im}a|$ on the right axis. The inset enlarges the hybridization region.}
\label{fig:hydrodynamic_breakdown}
\end{figure*}

Temperature and wave vector limit the hydrodynamic reduction through two distinct microscopic mechanisms: increasing temperature disrupts the isolated momentum-organized collision sector, whereas increasing wave vector couples the selected momentum-organized mode to other kinetic modes, reorganizing the constitutive transport channels. Figure~\ref{fig:hydrodynamic_breakdown} identifies the momentum-organized sector of the full collision operator and tracks its evolution under both variations.

At 17~K, the odd-parity sector of the NaF collision operator at $q=0$ contains an isolated eigenmode triplet with nearly unit momentum overlap $H_\lambda$ and large participation ratios $P_\lambda$ [Fig.~\ref{fig:hydrodynamic_breakdown}(a)]. Here $H_\lambda$ measures alignment with crystal momentum, while $P_\lambda$ measures the effective number of participating phonon states (Appendix~\ref{app:GK}). The spectral separation, strong momentum overlap, and large participation ratios of this triplet identify a momentum-organized hydrodynamic sector in the computed collision spectrum. Its longitudinal member, denoted $h$, carries the dominant relaxational heat-current contribution in the long-wavelength limit. It provides the microscopic basis for the single-relaxational-channel G--K kernel, with first-principles parameters listed in Table~S1.

This momentum-organized sector changes rapidly with temperature. The mode selected by the largest momentum overlap at each temperature loses both momentum overlap and spectral separation from neighboring odd-parity modes over approximately 21--24~K [Fig.~\ref{fig:hydrodynamic_breakdown}(b)]. Its momentum overlap reaches $H_\lambda=0.5$ at $T^*=22.9$~K. Over the same interval, its contributions to the instantaneous constitutive weight $Z_L(0,0^+)$ and bulk thermal conductivity $\kappa$ decrease sharply [Fig.~\ref{fig:hydrodynamic_breakdown}(c)]. These concurrent changes identify a redistribution of momentum and heat-current weight from a single isolated mode into multiple collision eigenmodes. The loss of an isolated momentum-organized sector does not imply immediate momentum relaxation: up to about 35~K, momentum correlations remain long-lived but are distributed among several slow modes (Sec.~S2\,A).

A distinct limitation emerges at finite wave vector, even when the momentum-organized sector remains isolated at $q=0$. At 17~K, the longitudinal hydrodynamic pole initially follows the perturbative G--K dispersion, $a_h^{\rm GK}(q)=\Gamma_h+\mu q^2$ [Fig.~\ref{fig:hydrodynamic_breakdown}(d)]. The nonlocal correction $\mu q^2$ arises from streaming-induced coupling between the slow odd-parity momentum mode and faster-relaxing even-parity collision modes, which carry no longitudinal heat current at $q=0$ by parity. The coupling grows linearly with $q$ and remains perturbative while it is small compared with the gap between collision-mode relaxation rates. The G--K constitutive kernel agrees with the full kinetic result to within 1\% for $L\gtrsim L_{\rm GK}=23$~mm (Table~S2), but higher-order streaming effects cause increasing deviations at shorter periods even while the hydrodynamic pole remains dominant.

At $L_{\rm pair}\approx3$~mm, two real constitutive eigenvalues coalesce and become a complex-conjugate pair, $a_\pm(q)$ [inset of Fig.~\ref{fig:hydrodynamic_breakdown}(d)]. This non-Hermitian pole merger is the nonperturbative continuation of the same streaming-induced coupling responsible for the G--K nonlocal correction. The resulting pair retains substantial heat-current weight and becomes the dominant streaming channel. G--K already loses quantitative accuracy at longer periods, and its single-real-pole structure cannot represent this transformation for any choice of parameters. A minimal two-mode model illustrates analytically how the same coupling produces both the perturbative correction and the pole merger (Appendix~\ref{app:kernel}).

\subsection{Evolution of transport-channel structure}
\label{sec:evolution_17K}
\begin{figure*}[!t]
\centering
\includegraphics[width=\textwidth]{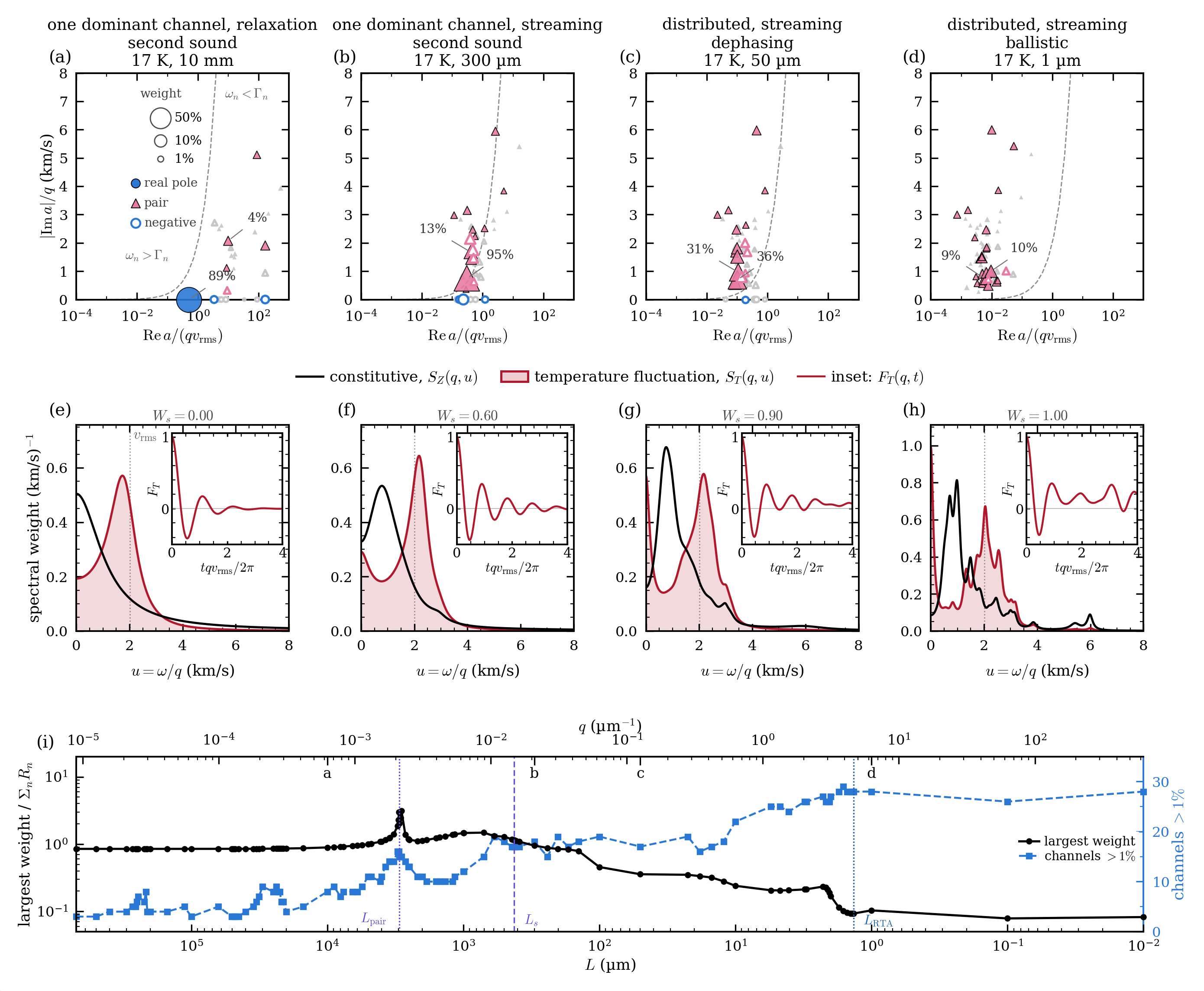}
\caption{\textbf{Evolution of transport-channel structure and temperature dynamics in NaF at 17~K.}
(a)--(d) Constitutive eigenvalues $a_n(q)$ and associated heat-current residues at $L=10$~mm, $300~\mu$m, $50~\mu$m, and $1~\mu$m, respectively. Circles denote real relaxational poles and triangles denote complex-conjugate streaming pairs. Marker sizes give the channel weights $w_n$ (Sec.~\ref{sec:diagnostics}) divided by $\sum_nR_n$. Open markers indicate negative signed weights, gray markers indicate weights below 1\%, and channels with weights below 0.2\% are omitted. Numbers label the signed weights of the two largest channels; dashed lines indicate $\omega_n=\Gamma_n$.
(e)--(h) Corresponding normalized constitutive spectra $S_Z(q,u)$ (black) and equilibrium temperature-fluctuation spectra $S_T(q,u)$ (red), displayed with a broadening of 0.1~km/s. Dotted lines mark $v_{\rm rms}$, insets show normalized temperature correlations $F_T(q,t)$, and values above the panels give the streaming index $W_s$.
(i) Largest channel weight, using the signed-sum normalization of (a)--(d), and number of channels with weights exceeding 1\%, as functions of grating period. Letters identify the four representative periods; vertical lines mark the onset of the streaming pair $L_{\rm pair}$, streaming crossover $L_s$, and RTA validity boundary $L_{\rm RTA}$. The largest weight peaks near $L_{\rm pair}$, where two real poles approaching coalescence carry large residues of opposite sign. Panel titles give the transport classes of Sec.~\ref{sec:diagnostics}.}
\label{fig:17K_two_levels}
\end{figure*}

Figure~\ref{fig:17K_two_levels} follows the transport channels, constitutive spectra, and temperature spectra at four representative periods at 17~K. Although the temperature response reverses sign at all four periods, the underlying transport-channel structure changes fundamentally with decreasing period: from a dominant relaxational channel to a dominant streaming pair, then to distributed streaming channels approaching independently damped phonon streaming. These examples illustrate how qualitatively similar oscillatory temperature dynamics can arise from distinct microscopic heat-current mechanisms.

\emph{Dominant relaxational channel.} At $L=10$~mm, a single real constitutive pole carries most of the signed heat-current weight [Fig.~\ref{fig:17K_two_levels}(a)], and the constitutive spectrum $S_Z$ is concentrated near zero frequency [Fig.~\ref{fig:17K_two_levels}(e)]. Nevertheless, the temperature-fluctuation spectrum $S_T$ exhibits a finite-frequency peak and the temperature correlation $F_T$ oscillates. This behavior is consistent with the G--K mechanism, in which energy-conservation feedback produces oscillatory temperature dynamics from a relaxational constitutive channel [Eq.~\eqref{eq:GK_FT_main}].

\emph{Dominant streaming channel.} At $L=300~\mu$m, the complex-conjugate streaming pair descended from the long-wavelength momentum-organized channel carries nearly all the signed heat-current weight [Fig.~\ref{fig:17K_two_levels}(b)]. The constitutive spectrum $S_Z$ develops a finite-frequency peak associated with this pair [Fig.~\ref{fig:17K_two_levels}(f)], while $S_T$ exhibits both a central relaxational component and a distinct finite-frequency peak. The temperature correlation $F_T$ remains strongly oscillatory. This period lies outside the oscillatory window of the G--K reduction. The light-scattering measurement at this period is discussed in Sec.~\ref{sec:discussion_temperature}.

\emph{Distributed collective streaming.} At $L=50~\mu$m, heat-current weight is redistributed among several streaming channels, with the two largest carrying approximately one-third of the total signed weight each [Fig.~\ref{fig:17K_two_levels}(c)]. The constitutive spectrum $S_Z$ develops additional finite-frequency structure, including higher-frequency contributions with smaller weights [Fig.~\ref{fig:17K_two_levels}(g)]. The temperature spectrum $S_T$ likewise exhibits multiple finite-frequency features, and $F_T$ remains oscillatory but departs from a single damped harmonic oscillation. Thus, temperature oscillations persist even after the dominant streaming pair gives way to distributed transport.

\emph{Streaming approaching RTA.} At still shorter periods, the relative influence of collision-induced coupling weakens, while heat-current weight remains distributed among many streaming channels. At $L=1~\mu$m, the largest channel carries only a small fraction of the total signed weight [Fig.~\ref{fig:17K_two_levels}(d)], and the full constitutive kernel approaches the energy-conserving RTA result. This period lies below the RTA validity boundary $L_{\rm RTA}$ and the collisionless crossover scale $L_{\rm cl}$ [Fig.~\ref{fig:17K_two_levels}(i) and Table~S2]. Both $S_Z$ and $S_T$ are broad, and $F_T$ becomes increasingly nonexponential, retaining oscillatory components associated with the superposition and dephasing of independently streaming phonons [Fig.~\ref{fig:17K_two_levels}(h)]. The onset of streaming-dominated transport occurs at periods far longer than the RTA validity boundary.

\subsection{Transport without an isolated momentum sector}
\label{sec:nonhydrodynamic}

\begin{figure*}[!t]
\centering
\includegraphics[width=\textwidth]{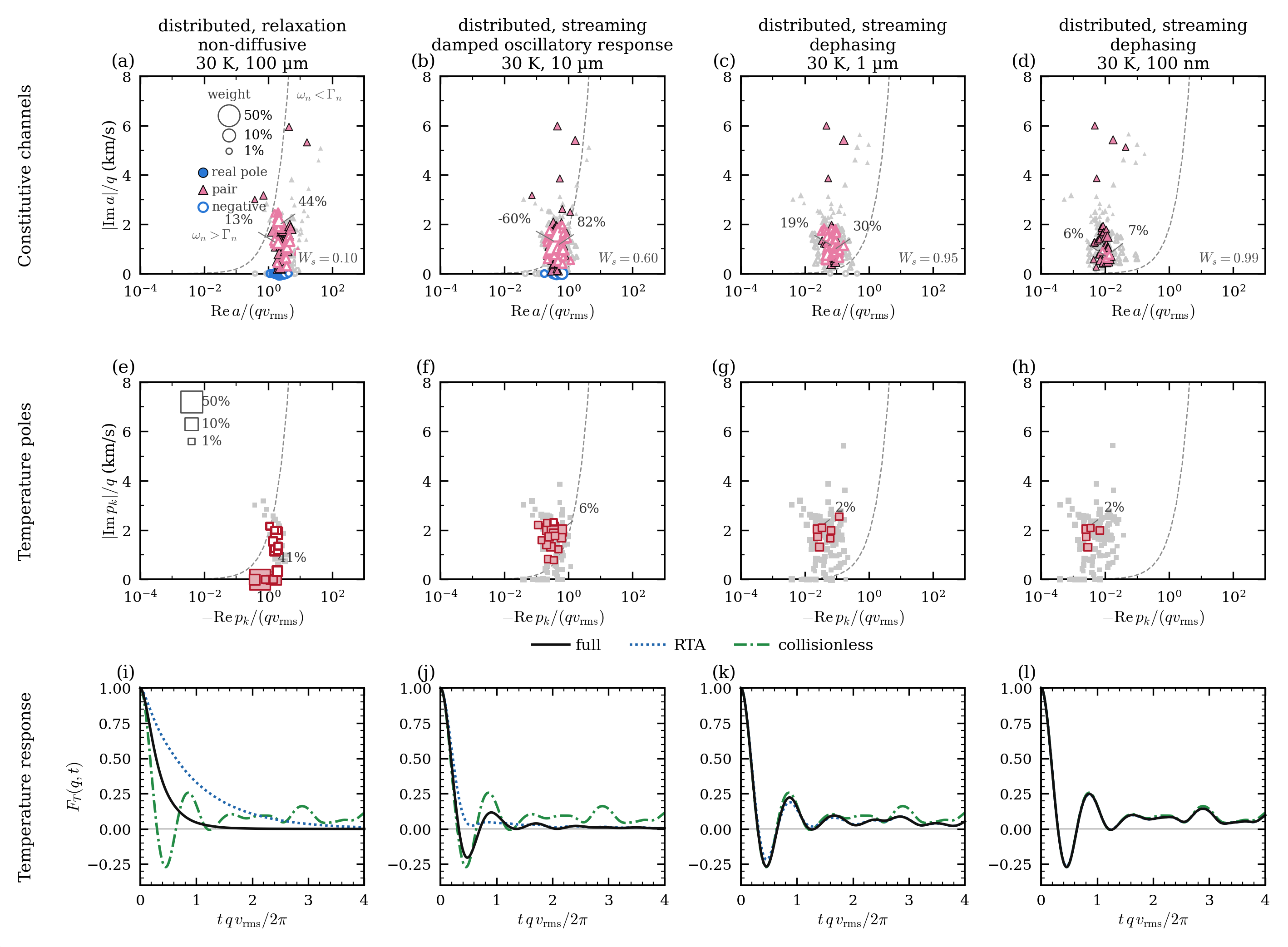}
\caption{\textbf{Collective relaxation, streaming, and the approach to RTA in NaF at 30~K.} (a)--(d) Constitutive eigenvalues $a_n(q)$ and heat-current residues at grating periods $L=100~\mu$m, $10~\mu$m, $1~\mu$m, and $100$~nm. Circles denote real poles and triangles conjugate pairs. Marker sizes give $w_n/\sum_nR_n$, open markers indicate negative signed weights, gray markers weights below 1\%, and dashed lines $\omega_n=\Gamma_n$. Numbers label the largest contributions. The streaming index $W_s$ is indicated in each panel. (e)--(h) Corresponding temperature poles $p_k(q)$, with marker sizes encoding their shares of $\sum_k|r_k|$ (conjugate partners combined) according to the size key. The horizontal coordinate is $-\operatorname{Re}\,p_k/(qv_{\rm rms})$, and the vertical coordinate is $|\operatorname{Im}\,p_k|/q$. Open squares indicate a residue with negative real part, gray squares shares below 1\%, and dashed curves $|\operatorname{Im}\,p_k|=|\operatorname{Re}\,p_k|$. (i)--(l) Normalized equilibrium temperature correlations $F_T(q,t)$ obtained from the full collision operator (solid black), RTA (dotted blue), and collisionless streaming (dash-dotted green). Time is normalized by $2\pi/(qv_{\rm rms})$. Panel titles give the transport classes of Sec.~\ref{sec:diagnostics}.}
\label{fig:30K_nonhydrodynamic}
\end{figure*}

\begin{figure*}[!t]
\centering
\includegraphics[width=\textwidth]{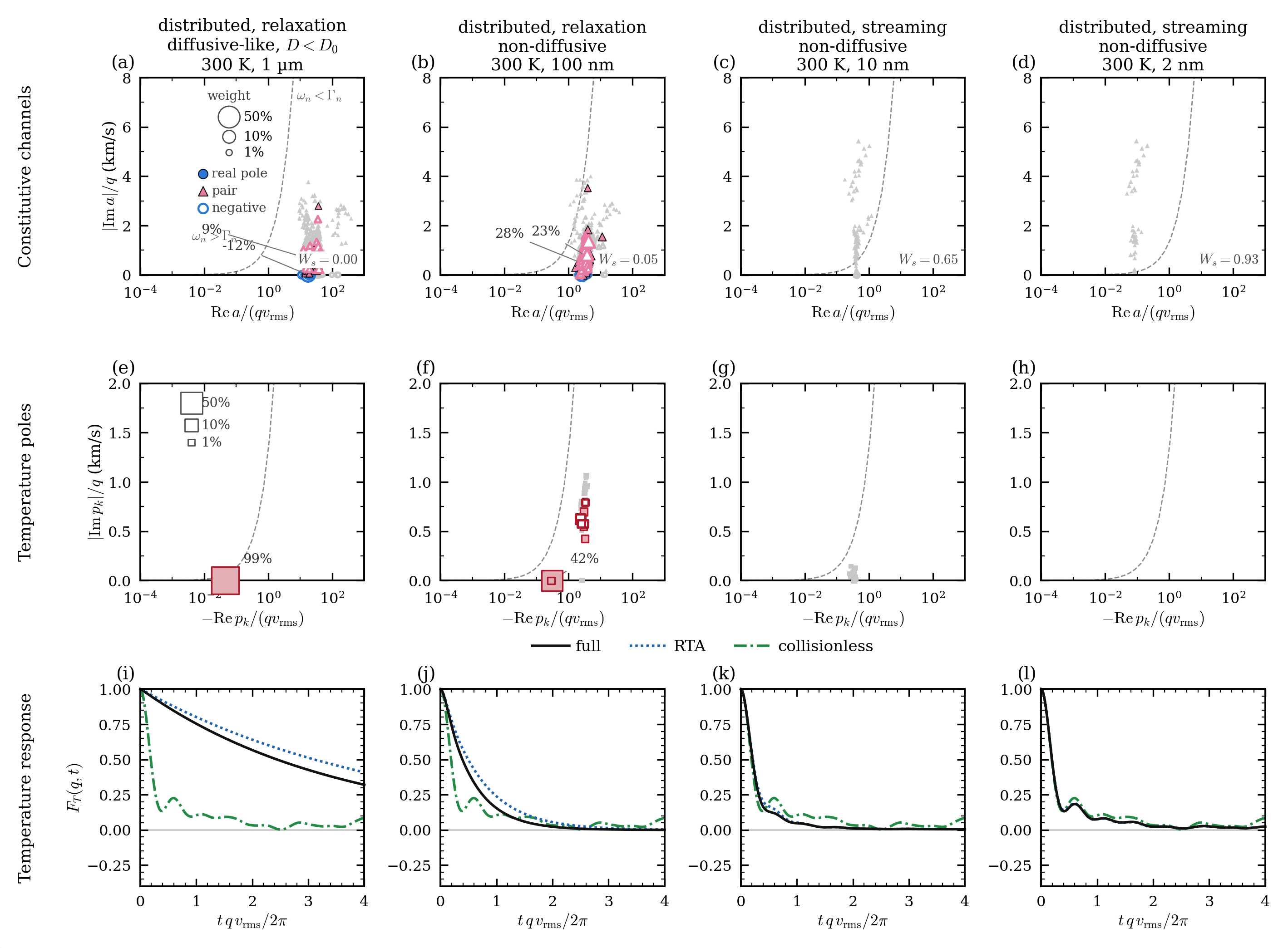}
\caption{\textbf{Constitutive streaming without temperature sign reversal in NaF at 300~K.} (a)--(d) Constitutive eigenvalues $a_n(q)$ and heat-current residues at grating periods $L=1~\mu$m, $100$~nm, $10$~nm, and $2$~nm, respectively. Circles denote real poles and triangles conjugate pairs. Marker sizes give $w_n/\sum_nR_n$, open markers indicate negative signed weights, gray markers weights below 1\%, and dashed lines $\omega_n=\Gamma_n$. Numbers label the largest contributions. (e)--(h) Corresponding temperature poles $p_k(q)$, plotted as $-\operatorname{Re}p_k/(qv_{\rm rms})$ versus $|\operatorname{Im}p_k|/q$ up to 2~km/s and sized by their shares of $\sum_k|r_k|$. Open squares indicate a residue with negative real part, gray squares shares below 1\%, and dashed curves $|\operatorname{Im}p_k|=|\operatorname{Re}p_k|$. At 10 and 2~nm the temperature weight is spread over many poles, each below 1\% (gray) or below the 0.2\% display cutoff. (i)--(l) Normalized equilibrium temperature correlations from the full collision operator (solid black), RTA (dotted blue), and collisionless streaming (dash-dotted green). Time is normalized by $2\pi/(qv_{\rm rms})$. Panel titles give the transport classes of Sec.~\ref{sec:diagnostics}.}
\label{fig:300K_nonhydrodynamic}
\end{figure*}

Above $T^*$, the collision operator has no isolated momentum sector, which removes the basis for the isolated-channel hydrodynamic reduction of Secs.~\ref{sec:hydrodynamic_breakdown} and \ref{sec:evolution_17K}. Figures~\ref{fig:30K_nonhydrodynamic} and \ref{fig:300K_nonhydrodynamic} follow the two spectral levels at 30 and 300~K over four grating periods each. They show that collective collision coupling remains strong without an isolated momentum sector, and that the same constitutive progression produces opposite temperature behavior at the two temperatures.

\emph{Collective thermal wave at 25~K.} Between the loss of the isolated sector and the distributed regime at 30~K, the leading temperature-pole pair is still a distinguishable collective thermal wave: over 85--$250~\mu$m it carries a substantial share of the temperature weight, is underdamped, has a narrow frequency distribution and disperses continuously with wave vector, although the G--K oscillatory window closes at 21~K (Secs.~S2).

\emph{Distributed channels at 30~K.} Collision coupling among phonons persists after the isolated momentum sector is lost, with the heat-current response carried by several channels. At 30~K and $L=100~\mu$m, the heat-current weight is distributed among several overdamped channels [Fig.~\ref{fig:30K_nonhydrodynamic}(a)]. The temperature correlation relaxes without sign reversal through several modes, whereas the RTA correlation decays more slowly [Fig.~\ref{fig:30K_nonhydrodynamic}(e,i)]. At $L=10~\mu$m the kernel becomes streaming-dominated through channels whose signed weights partly cancel, and $F_T$ reaches a minimum of $-0.20$ followed by a positive lobe that RTA does not reproduce [Fig.~\ref{fig:30K_nonhydrodynamic}(b,f,j)]. Collective streaming and oscillatory temperature dynamics arise from distributed channels at 30~K.

\emph{Approach to RTA.} With decreasing period, streaming grows at 30 and 300~K [Figs.~\ref{fig:30K_nonhydrodynamic}(c,d) and \ref{fig:300K_nonhydrodynamic}(a)--(d)], and the full kernel approaches the RTA kernel, reaching it within 1\% below $L_{\rm RTA}$ [Table~S2]. The streaming crossover lies at much shorter periods at 300~K than at 30~K, because the streaming frequencies $qv_{\alpha L}$ must exceed correspondingly larger collision rates. As the response approaches RTA, heat-current weight spreads over many streaming channels, and the temperature poles span a broad range of frequencies and decay rates. The full and RTA temperature correlations converge [Figs.~\ref{fig:30K_nonhydrodynamic}(g,h,k,l) and \ref{fig:300K_nonhydrodynamic}(g,h,k,l)].

\emph{Streaming without temperature sign reversal.} The temperature correlation at 30~K reverses sign below about $50~\mu$m (Table~S2), whereas at 300~K it remains positive at every period [Fig.~\ref{fig:300K_nonhydrodynamic}(i)--(l)]. The difference persists in the collisionless limit, whose first minimum crosses zero near 70~K. By Eq.~\eqref{eq:collisionless_FT_main}, this follows from the heat-capacity weighting of the streaming velocities: at 300~K a larger share of the heat capacity lies in phonons with small projected velocities (Sec.~S2\,C). These phonons contribute to temperature fluctuations through their heat capacities, but their longitudinal heat-current weight is suppressed by the additional factor $v_{\alpha L}^2$. Therefore, strong constitutive streaming does not guarantee temperature sign reversal, even when collisions are absent.

\section{Discussion}
\label{sec:discussion}
Transport channels, the accuracy of reduced constitutive models, and the temperature response change at different scales. The constitutive spectrum separates the first from the other two.

\subsection{Hidden transport channels}
\label{sec:discussion_reductions}
The constitutive and temperature spectra reveal distinct aspects of heat conduction: the former identifies the microscopic channels carrying heat current, whereas the latter describes their combined dynamics coupled to conserved energy. This coupling can produce temperature behavior that does not directly reflect the character of the underlying constitutive channels. A relaxational heat-current channel can generate oscillatory temperature dynamics through energy-conservation feedback, as at 10~mm and 17~K [Fig.~\ref{fig:17K_two_levels}(a,e)], whereas streaming-dominated transport can produce temperature decay without sign reversal, as at 300~K. Oscillatory temperature dynamics can also arise from a dominant streaming pair or distributed streaming channels (Fig.~\ref{fig:17K_two_levels}). Thus, similar observable temperature behavior can conceal fundamentally different organizations of microscopic heat-current transport.

Previous first-principles approaches share the microscopic foundation of the present theory, namely the linearized phonon Boltzmann equation with the full collision operator, but resolve different aspects of thermal transport. The transport waves introduced by Cepellotti and Marzari are eigenmodes of the full kinetic generator, including the conserved energy coordinate~\cite{cepellotti2017transport}. Modes with nonzero temperature-response residues contribute to the temperature spectrum, but an oscillatory transport wave does not necessarily represent an oscillatory constitutive heat-current channel. Similarly, the full-scattering-matrix Green’s functions developed by Chiloyan \emph{et al.} describe phonon-population responses and the resulting temperature dynamics~\cite{chiloyan2021green}. The generalized conductivity of Hua and Lindsay captures the aggregate transient and nonlocal heat-current response through a kernel corresponding to $\tilde Z_L$, without resolving its individual transport channels~\cite{hua2020space}. In contrast, our theory explicitly resolves the constitutive response into heat-current transport channels characterized by their poles and residues, revealing how the microscopic organization of transport evolves through channel hybridization and spectral-weight redistribution. By connecting this constitutive spectrum to the distinct spectrum of temperature fluctuations through energy conservation, the theory further establishes why similar observable temperature dynamics can arise from fundamentally different heat-current transport mechanisms.

This additional spectral level provides a microscopic criterion for assessing reduced constitutive theories beyond their agreement with observable temperature dynamics. In NaF, the G--K reduction loses quantitative accuracy before the dominant relaxational channel transforms into a streaming pair, whose complex-conjugate poles cannot be represented by the single-real-pole constitutive structure (Sec.~\ref{sec:hydrodynamic_breakdown}). The pair nevertheless retains substantial momentum-derived heat-current weight, while the full temperature response remains oscillatory below the short-period edge of the G--K oscillatory window. The channel analysis thus distinguishes the failure of a reduced constitutive description from the disappearance of the underlying transport mechanism. More generally, comparing the constitutive kernels and their dominant poles and residues reveals whether a reduced theory faithfully represents the heat-current dynamics over the relevant temperature, wave-vector, and frequency ranges.

\subsection{Physical origins of transport crossovers}
\label{sec:map}
\begin{figure*}[!t]
\centering
\includegraphics[width=\textwidth]{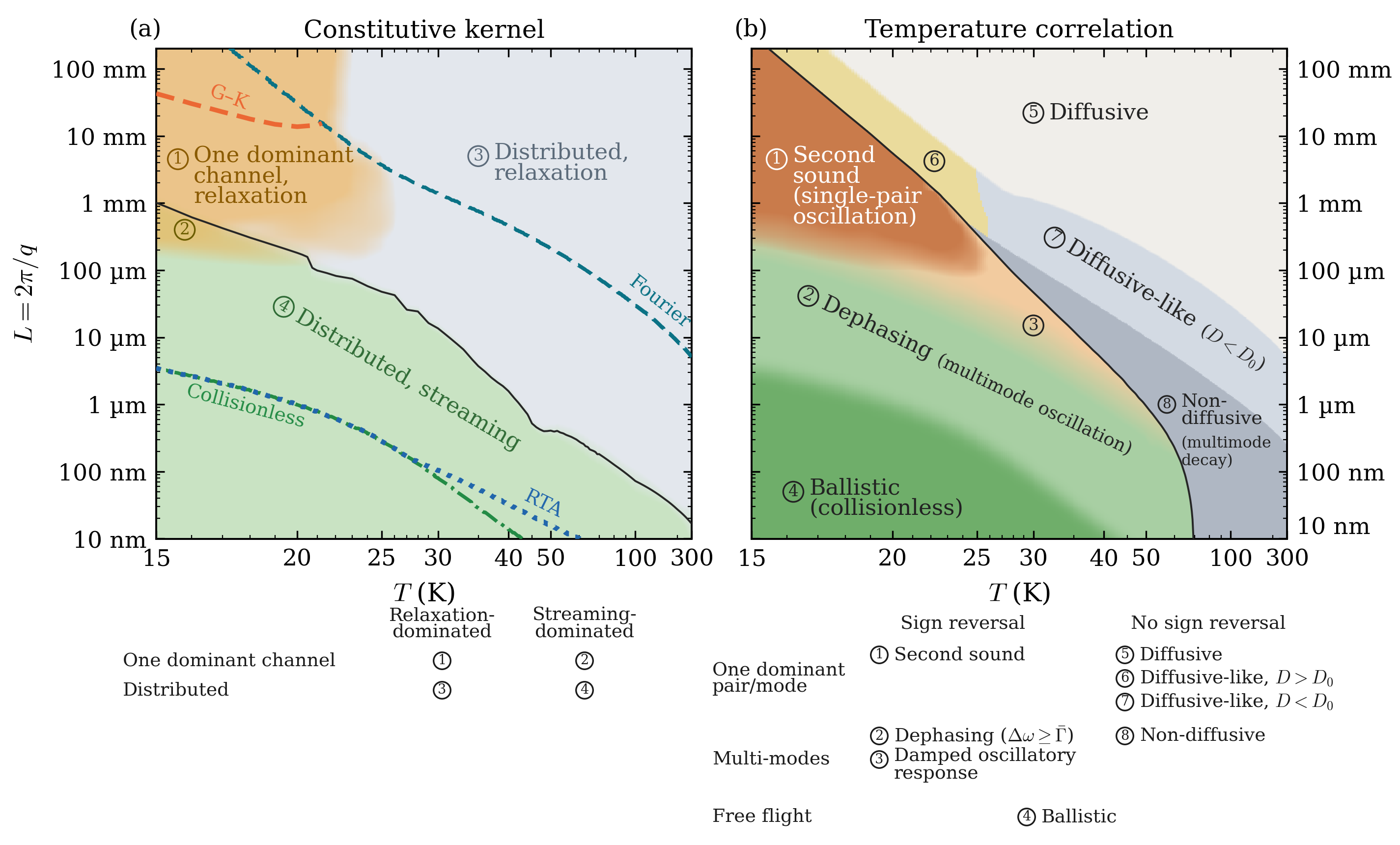}
\caption{\textbf{Two maps of heat conduction in NaF.} The maps are plotted against temperature (axis linear in $1/T$) and spatial period $L$. Circled numbers refer to the class keys below the panels. (a) Constitutive kernel, classified by the largest grouped channel share $\hat w_1$ and the streaming index $W_s$, each at threshold 1/2. The black line is the streaming crossover $L_s$. Colored lines indicate operational validity boundaries, defined by a 1\% deviation of each reduced constitutive kernel from the full kernel (Appendix~\ref{app:boundaries}). Fourier holds above $L_F$ (teal dashed), and G--K above $L_{\rm GK}$ (orange dashed, shown only where $L_{\rm GK}<L_F$). The blue dotted line marks the onset of validity of RTA or its collisionless limit: it follows $L_{\rm RTA}$ except where $L_{\rm cl}>L_{\rm RTA}$ (below 27.5~K), where it follows $L_{\rm cl}$. Collisionless streaming holds below $L_{\rm cl}$ (green dash-dot). (b) Temperature correlation, classified by sign reversal of $F_T$ (black line), the leading-pair share $P_{\rm pair}$ (threshold 1/2), the dephasing ratio $\Delta\omega/\bar\Gamma$ (threshold 1), and, without sign reversal, the decay amplitude $P_D$ and $D(q)/D_0$. Ballistic cases lie below $L_{\rm cl}$ of panel (a). Shading blends across each threshold. Definitions are in Sec.~\ref{sec:diagnostics}, and line values at selected temperatures in Table~S2.}
\label{fig:maps}
\end{figure*}

Wave vector and temperature limit the hydrodynamic reduction through different physical changes. Increasing wave vector leaves the collision spectrum unchanged but strengthens the streaming coupling between its modes. The same coupling first produces the G--K nonlocal correction and then the pole merger. Increasing temperature instead changes the collision spectrum itself: it redistributes momentum and heat-current weight among collision modes and weakens the spectral isolation on which the single-channel reduction relies. Momentum remains long-lived after it is no longer held by one isolated mode, and the collective thermal wave at 25~K shows that collective propagation also persists beyond that sector. Collective transport encompasses more than the isolated-sector hydrodynamic reduction.

Streaming can dominate the heat-current response while collision coupling still shapes its channels. RTA omits the off-diagonal collision couplings and the collisionless limit omits all collisions, so their accuracy measures how much these couplings change the kernel. In NaF, each reduced description lies almost entirely within one kernel class of Fig.~\ref{fig:maps}(a), and none covers the dominant-streaming-channel class. Streaming dominates the kernel at periods far longer than those at which RTA and collisionless streaming hold. In between, off-diagonal couplings continue to set the frequencies and heat-current residues of the streaming channels. At low temperature, collision-induced redistribution preserves much of the heat current, so its net relaxation is substantially slower than phonon out-scattering. RTA omits the collision coupling responsible for this current-preserving redistribution, so at these temperatures the full kernel reaches the collisionless tolerance before the RTA tolerance, and the two validity boundaries follow no universal order.

The temperature map does not follow the kernel map, because energy conservation combines the channels into temperature modes and the two levels weight phonons differently. The phonon weights are $c_\alpha$ for temperature fluctuations and $c_\alpha v_{\alpha L}^2$ for the instantaneous current response, so at 300~K slow phonons suppress temperature sign reversal even in the collisionless reference while the kernel remains streaming-dominated. The sign-reversal line of Fig.~\ref{fig:maps}(b) does not follow the streaming crossover of Fig.~\ref{fig:maps}(a). Memory and spatial nonlocality can also offset one another in the temperature decay rate. In the slow-pole approximation, spatial nonlocality suppresses the static conductivity and slows temperature relaxation, whereas memory can increase its decay rate (Appendix~\ref{app:fourier}). Their partial cancellation can leave the decay rate close to the Fourier value while the kernel retains both corrections.

\subsection{Implications for temperature measurements}
\label{sec:discussion_temperature}
A complete, ideal temperature response determines the aggregate constitutive kernel in principle, because energy conservation links the two exactly. Finite bandwidth and noise make the extraction of its poles and residues ill-conditioned. Resonance positions, oscillations, and sign reversals constrain the channel organization only partly, because each can arise from different channel organizations. Likewise, an effective conductivity fitted with a diffusive model samples the frequency-dependent kernel and generally differs from the static nonlocal conductivity $\kappa(q)=\tilde Z_L(q,0)$. Comparison with transient-grating measurements assumes a signal proportional to the temperature modulation and a phonon population initially in local equilibrium.

Wave-vector scans provide direct tests of the contrasting temperature dynamics predicted by the full microscopic theory and reduced constitutive descriptions. At 17~K, the independently parameterized G--K reduction predicts a return to overdamped temperature dynamics below the short-period edge of its oscillatory window, about 0.76~mm, whereas the full theory predicts persistent oscillations at shorter periods. A systematic grating-period scan across this boundary at fixed temperature would directly test this qualitative difference. The predicted dispersion provides an additional distinction: the apparent propagation speed increases toward shorter periods in the full theory as the response approaches collisionless dephasing, whereas the extrapolated G--K response predicts a decreasing speed (Fig.~S4(a)). Further predictions accessible to transient-grating measurements include oscillatory temperature dynamics at 25~K over 85--$250~\mu$m and temperature sign reversal at 30~K over 10--$30~\mu$m, where the full temperature correlation becomes negative while RTA remains positive (Sec.~S2\,C). Phase-sensitive measurements could resolve these contrasting responses.

The historical NaF light-scattering experiment of Pohl and Irniger is consistent with this qualitative difference: a weakly damped thermal resonance was observed at $L=295~\mu$m over 17--21.5~K~\cite{pohl1976observation}, below the G--K short-period edge at 17~K. The full first-principles calculation also gives an oscillatory response and predicts a resonance speed close to the measured value at 17~K. In the calculation, the constitutive kernel at this period is dominated by the streaming pair formed from the momentum-organized relaxational channel. The calculated resonance is broader than observed, and the agreement in speed deteriorates at higher temperatures (Sec.~S1\,B). The experimental resonance can be fitted with a G--K-type temperature response using adjustable damping, but such a fit does not establish the validity of its underlying single-relaxational-channel constitutive description.

\section{Conclusions}
\label{sec:conclusion}
Behind observable temperature dynamics lies a hidden organization of heat-current transport channels that can change fundamentally without a corresponding qualitative change in temperature behavior. Oscillatory temperature dynamics can arise from relaxational constitutive channels through energy-conservation feedback, while streaming-dominated transport need not produce temperature sign reversal. These findings demonstrate that observable temperature oscillations or diffusion-like decay alone cannot uniquely identify the microscopic mechanisms governing heat conduction. By resolving the relaxation and streaming channels, their hybridization, and the redistribution of heat-current weight, the constitutive spectrum reveals the physical organization of non-Fourier heat conduction that remains hidden in aggregate conductivity and temperature measurements.

First-principles calculations for NaF demonstrate two distinct routes beyond phonon hydrodynamics. At low temperature, increasing wave vector drives the isolated momentum-organized relaxational channel into a complex-conjugate streaming pair through the same mode coupling responsible for the perturbative G--K nonlocal correction. This nonperturbative transformation invalidates the single-relaxational-channel constitutive structure, while oscillatory temperature dynamics persist beyond the short-period edge of the G--K oscillatory window. Increasing temperature instead removes the spectral isolation of the momentum-organized sector, yet oscillatory temperature dynamics can survive through other collision-coupled channels. At shorter wavelengths, heat-current weight spreads among multiple collision-coupled streaming channels before approaching independently damped phonon streaming, revealing an intermediate organization of heat transport that is captured by neither the single-channel G--K reduction nor independent-phonon descriptions.

The two spectral levels therefore provide complementary descriptions of heat conduction. Agreement with an observed temperature response does not by itself validate the microscopic mechanism assumed by a reduced constitutive model. A physically faithful reduction must also reproduce the relevant constitutive response and its dominant transport-channel organization.

Wave-vector-dependent thermal measurements can test these predictions. Recent observations of thermal waves in isotopically enriched graphite~\cite{xie2026room} offer another setting in which to resolve the underlying transport channels. Second sound in ultracold Fermi and Bose gases~\cite{sidorenkov2013second,christodoulou2021observation,yan2024thermography}, where density and entropy dynamics are coupled, offers a setting for extending the channel analysis beyond phonon heat conduction. The separation of constitutive channels from conserved-density dynamics also motivates analogous analyses of charge, spin, and mass transport.

\section*{Data and code availability}

Data supporting this study, and the code used for the calculations and figures, are available from the corresponding authors on reasonable request.

\begin{acknowledgments}
This work was authored in part by the National Laboratory of the Rockies for the U.S. Department of Energy (DOE), operated under Contract No. DE-AC36-08GO28308. Funding provided by the U.S. Department of Energy Office of Science Energy Earthshots Initiative as part of the Degradation Reactions in Electrothermal Energy Storage (DEGREES) project. The views expressed in the article do not necessarily represent the views of the DOE or the U.S. Government. The U.S. Government retains and the publisher, by accepting the article for publication, acknowledges that the U.S. Government retains a nonexclusive, paid-up, irrevocable, worldwide license to publish or reproduce the published form of this work, or allow others to do so, for U.S. Government purposes. This research was performed using computational resources sponsored by the U.S. Department of Energy's Office of Critical Minerals and Energy Innovation and located at the National Laboratory of the Rockies. J.D. acknowledges support from the Thomas and Jean Walter endowment of Auburn University. We thank D.~G.~Cahill for discussions on the selection of candidate materials for the calculations. The authors used Claude (Opus 5.5, Anthropic) and Codex (GPT-6.1 Sol, OpenAI) for manuscript polishing. The authors verified the outputs and take full responsibility for the article.
\end{acknowledgments}

\appendix

\section{Projected phonon kinetics and constitutive spectrum}
\label{app:projected}

\subsection{Symmetrized representation and streaming vertices}
\label{app:methods_projected_transport}

A phonon state $\alpha\equiv(\mathbf k,b)$ has frequency $\omega_\alpha$, group velocity $\mathbf v_\alpha=\nabla_{\mathbf k}\omega_\alpha$, and mode heat capacity
\begin{equation}
c_\alpha=\hbar\omega_\alpha\frac{\partial n_\alpha^0}{\partial T},
\qquad C=\sum_\alpha c_\alpha,
\qquad c_V=\frac{C}{V},
\label{eq:mode_heat_capacity_appendix}
\end{equation}
where $n_\alpha^0$ is the equilibrium Bose occupation. We use the symmetrized, heat-capacity-weighted representation of the linearized phonon Boltzmann equation~\cite{zeng_and_dong_2026_unified}. The collision operator $\boldsymbol\Omega$ is real symmetric and positive semidefinite, while the velocity operators $\mathbf V$ are diagonal in the phonon basis.

The energy null vector $|e\rangle$ is even under parity, with components $e_\alpha=(c_\alpha/C)^{1/2}$.
We use the full generator $\boldsymbol{\mathcal K}$ and energy-orthogonal projector $\mathbf Q=\mathbf I-|e\rangle\langle e|$ defined in Sec.~\ref{sec:constitutive_spectrum}, with projected generator $\boldsymbol\Xi$ given by Eq.~\eqref{eq:transport_generator}. The full generator governs the coupled evolution of energy and nonconserved phonon populations, whereas the projected generator determines the constitutive response to a prescribed temperature field.

At $q=0$, the collision eigenmodes can be classified by inversion parity. We denote the odd modes by $|\chi_\lambda\rangle$, with collision rates $\gamma_\lambda$, and the even modes orthogonal to $|e\rangle$ by $|\nu_\iota\rangle$, with rates $\varepsilon_\iota$. In the energy-plus-collision basis, the kinetic state is
\begin{equation}
|\Psi\rangle=
\begin{pmatrix}
\Theta\\
\boldsymbol\eta\\
\boldsymbol\zeta
\end{pmatrix},
\qquad
\Theta=\left(\frac{C}{k_{\rm B}T_0^2}\right)^{1/2}\delta T,
\label{eq:projected_state_appendix}
\end{equation}
where $\boldsymbol\eta$ and $\boldsymbol\zeta$ contain the odd- and even-sector amplitudes. The streaming vertices are
\begin{subequations}
\label{eq:streaming_vertices_appendix}
\begin{align}
\boldsymbol\Lambda_\lambda &=\sum_\alpha\left(\frac{c_\alpha}{C}\right)^{1/2}\chi_{\lambda\alpha}\mathbf v_\alpha, \label{eq:Lambda_vertex_appendix}\\
\boldsymbol\Pi_{\lambda\iota} &=\sum_\alpha\chi_{\lambda\alpha}\nu_{\iota\alpha}\mathbf v_\alpha. \label{eq:Pi_vertex_appendix}
\end{align}
\end{subequations}
Because velocity is odd under inversion, streaming couples opposite-parity sectors. For $\mathbf q=q\hat x$, the projected generator in the longitudinal symmetry sector takes the form
\begin{equation}
\boldsymbol\Xi(q)=
\begin{pmatrix}
\boldsymbol\gamma&iq\boldsymbol\Pi_x\\
iq\boldsymbol\Pi_x^{\mathsf T}&\boldsymbol\varepsilon
\end{pmatrix},
\label{eq:Xi_methods_short}
\end{equation}
where $\boldsymbol\gamma$ and $\boldsymbol\varepsilon$ are diagonal collision-rate matrices. Here $\boldsymbol\Lambda_\lambda$ and $\boldsymbol\Pi_{\lambda\iota}$ are physical-space vectors, whereas $\boldsymbol\Lambda_x$ is the column of components $(\boldsymbol\Lambda_x)_\lambda=\Lambda_{\lambda x}=\langle\chi_\lambda|\Lambda_x\rangle$ over the odd modes and $\boldsymbol\Pi_x$ is the odd--even coupling matrix $(\boldsymbol\Pi_x)_{\lambda\iota}=\Pi_{\lambda\iota,x}$. The longitudinal current vertex $\mathbf g$ is the representation of $|\Lambda_x\rangle$ from Sec.~\ref{sec:constitutive_spectrum} in this collision basis, $\mathbf g=(\boldsymbol\Lambda_x,\mathbf 0)^{\mathsf T}$. The full generator becomes
\begin{equation}
\boldsymbol{\mathcal K}(q)=
\begin{pmatrix}
0&iq\mathbf g^{\mathsf T}\\
iq\mathbf g&\boldsymbol\Xi(q)
\end{pmatrix}.
\label{eq:E_methods_appendix}
\end{equation}
The transposes are unconjugated. The off-diagonal blocks couple the conserved energy coordinate to the projected kinetic sector and generate the feedback responsible for temperature evolution.

\subsection{Constitutive kernel from projected dynamics}
\label{app:kernel}

The heat current is
\begin{equation}
j_i=\frac{\sqrt{Ck_{\rm B}T_0^2}}{V}\sum_\lambda\Lambda_{\lambda i}\eta_\lambda.
\label{eq:projected_heat_current_appendix}
\end{equation}
For a prescribed temperature field, the projected amplitudes obey
\begin{equation}
\partial_t|\Psi_Q\rangle=-\boldsymbol\Xi|\Psi_Q\rangle-iq\mathbf g\Theta.
\label{eq:projected_driven_appendix}
\end{equation}
With vanishing initial kinetic deviation, Laplace transformation gives $|\tilde\Psi_Q\rangle=-iq(s\mathbf I+\boldsymbol\Xi)^{-1}\mathbf g\tilde\Theta$. Substitution into the current expression yields
\begin{equation}
\tilde Z_L(q,s)=c_V\mathbf g^{\mathsf T}[s\mathbf I+\boldsymbol\Xi(q)]^{-1}\mathbf g.
\label{eq:projected_kernel_derivation_appendix}
\end{equation}
More generally, $\tilde Z_{ij}(\mathbf q,s)=c_V\langle\Lambda_i|[s\mathbf I+\boldsymbol\Xi(\mathbf q)]^{-1}|\Lambda_j\rangle$. This establishes the constitutive resolvent in Eq.~\eqref{eq:constitutive_spectrum_general} without imposing energy-conservation feedback on the prescribed temperature field.

For diagonalizable $\boldsymbol\Xi$, the right and left eigenvectors form a biorthogonal basis satisfying $\sum_n|\phi_n^{\mathrm R}\rangle\langle\phi_n^{\mathrm L}|=\mathbf I$ on $\operatorname{Ran}\mathbf Q$. Inserting this completeness relation into the constitutive resolvent gives the pole-residue expansion of Eq.~\eqref{eq:constitutive_spectrum_general}.
Only modes with nonzero current residues contribute to the constitutive response. The eigenvalues determine their relaxation and streaming frequencies, while the residues determine their coupling to the heat current. At defective points, the resolvent must instead be represented using the corresponding invariant subspaces or Jordan blocks.

\emph{Two-mode illustration of streaming-induced hybridization.} The connection between the perturbative G--K correction and the nonperturbative transformation of the constitutive channel can be illustrated by restricting the longitudinal generator [Eq.~\eqref{eq:Xi_methods_short}] to the hydrodynamic odd mode $h$ and one even mode $\iota$, with collision rates $\Gamma_h<\varepsilon_\iota$ and streaming vertex $\Pi\equiv\hat{\mathbf q}\cdot\boldsymbol\Pi_{h\iota}$:
\begin{equation}
\boldsymbol\Xi_2(q)=
\begin{pmatrix}
\Gamma_h & iq\Pi\\
iq\Pi & \varepsilon_\iota
\end{pmatrix},
\label{eq:two_mode_generator}
\end{equation}
with eigenvalues
\begin{equation}
a_\pm(q)=\frac{\Gamma_h+\varepsilon_\iota}{2}\pm\sqrt{\frac{(\varepsilon_\iota-\Gamma_h)^2}{4}-q^2\Pi^2}.
\label{eq:two_mode_eigenvalues}
\end{equation}
At small wave vector the slow branch is $a_h(q)=\Gamma_h+\mu_2q^2+O(q^4)$, with $\mu_2=\Pi^2/(\varepsilon_\iota-\Gamma_h)$ the single-mode term of the G--K coefficient $\mu$ [Eq.~\eqref{eq:mu_methods_short}]. The same coupling makes the two real eigenvalues coalesce at $q_{\rm pair}=(\varepsilon_\iota-\Gamma_h)/(2|\Pi|)$, beyond which they form a complex-conjugate pair. With the heat-current vertex $\Lambda_h$ on the odd mode, the constitutive kernel is
\begin{equation}
\tilde Z_2(q,s)=c_V\Lambda_h^2\,\frac{s+\varepsilon_\iota}{(s+\Gamma_h)(s+\varepsilon_\iota)+q^2\Pi^2}.
\label{eq:two_mode_kernel}
\end{equation}
The individual residues diverge at the coalescence, while their sum stays finite. One streaming coupling thus produces both the perturbative nonlocal correction and, at larger wave vector, the change of channel structure. In NaF the hydrodynamic mode couples to many even modes, so the model is illustrative; evaluated with the $q=0$ rate and vertex of the even mode most strongly coupled to $h$, it nevertheless gives $2\pi/q_{\rm pair}$ close to the computed $L_{\rm pair}$ on the denser meshes (Sec.~S2\,A).

\subsection{Spectral identities and real-frequency representation}
\label{app:spectral}

\emph{Instantaneous-weight sum rules.} Biorthogonal completeness and inversion symmetry give
\begin{equation}
\begin{aligned}
\sum_nR_{ij}^{(n)}(\mathbf q) &=\lim_{s\to\infty}s\tilde Z_{ij}(\mathbf q,s) =c_V\langle\Lambda_i|\Lambda_j\rangle\\
& =c_V\langle e|\mathbf V_i\mathbf V_j|e\rangle =\frac{1}{V}\sum_\alpha c_\alpha v_{\alpha i}v_{\alpha j}.
\end{aligned}
\end{equation}
This sum is independent of $\mathbf q$. In phonon variables, $v_{\rm rms}^2=\sum_\alpha c_\alpha v_{\alpha L}^2/C$, where $v_{\alpha L}=\hat{\mathbf q}\cdot\mathbf v_\alpha$. The instantaneous-weight sum rule constrains the total response even when individual residues are signed or complex.

\emph{Conjugate-pair contributions.} For a conjugate pair $a_\pm=\Gamma\pm i\omega$ with residues $R_-=R_+^*$, the combined time-domain contribution is Eq.~\eqref{eq:streaming_channel_time}. Its instantaneous contribution is $2\operatorname{Re}R_+$, which may be negative even though the complete real-frequency constitutive response is nonnegative. Individual pole residues cannot generally be interpreted as positive fractions of transported heat.

\emph{Positivity and normalization.} For real $\omega$, let $\mathbf A=i\omega\mathbf I+\boldsymbol\Xi$. Its Hermitian part is $(\mathbf A+\mathbf A^\dagger)/2=\mathbf Q\boldsymbol\Omega\mathbf Q\geq0$. The current vertex $\mathbf g$ is real. Wherever $\mathbf A$ is invertible,
\begin{equation}
\operatorname{Re}(\mathbf g^\dagger\mathbf A^{-1}\mathbf g)=(\mathbf A^{-1}\mathbf g)^\dagger\boldsymbol\Omega(\mathbf A^{-1}\mathbf g)\geq0.
\label{eq:positive_real_appendix}
\end{equation}
Consequently, $\operatorname{Re}\tilde Z_L(q,i\omega)\geq0$. Causality and the instantaneous-weight sum rule imply
\begin{equation}
\int_{-\infty}^{\infty}\operatorname{Re}\tilde Z_L(q,i\omega)\,d\omega=\pi c_Vv_{\rm rms}^2.
\label{eq:frequency_sum_appendix}
\end{equation}
For an inversion-symmetric crystal, the integrand is even in frequency, and substituting $\omega=qu$ gives $\int_0^\infty S_Z(q,u)\,du=1$ for the representation of Eq.~\eqref{eq:constitutive_speed_spectrum}.
These relations hold for $q>0$. In collisionless or partially undamped systems, the real-frequency response may contain delta-function contributions, and the sum rule is understood distributionally. The representation $S_Z$ combines the constitutive poles and residues into a positive spectral profile, but its maxima need not coincide with individual constitutive eigenfrequencies.

\emph{Complex symmetry and pole coalescence.} The projected generator satisfies
\begin{equation}
\boldsymbol\Xi^{\mathsf T}(q)=\boldsymbol\Xi(q).
\label{eq:complex_symmetric_appendix}
\end{equation}
Away from defective points, writing $\boldsymbol\phi_n^{\mathrm R}$ for the right eigenvector $|\phi_n^{\mathrm R}\rangle$ in the kinetic coordinates, the longitudinal residue can be written using the unconjugated product as
\begin{equation}
R_n=c_V\frac{(\mathbf g^{\mathsf T}\boldsymbol\phi_n^{\mathrm R})^2}{(\boldsymbol\phi_n^{\mathrm R})^{\mathsf T}\boldsymbol\phi_n^{\mathrm R}}.
\label{eq:complex_symmetric_residue_appendix}
\end{equation}
The unconjugated product determines the response amplitude, whereas modal character is evaluated using Hermitian overlaps. Where two poles coalesce, the coalescing eigenvectors may become self-orthogonal, causing individual simple-pole residues to diverge. The combined resolvent remains well defined away from its poles, although a defective eigenvalue can generate higher-order poles. Therefore, we group coalescing channels and track their invariant subspaces. Numerical examples are given in Sec.~S2\,A.

\section{Equilibrium temperature-fluctuation spectrum}
\label{app:projected_current_memory}
\label{app:methods_temperature_response}

\subsection{Equilibrium covariance and temperature correlation}

In the semiclassical phonon-population description, define the normalized occupation fluctuations
\begin{equation}
\varphi_\alpha=\frac{\delta n_\alpha}{\sqrt{n_\alpha^0(1+n_\alpha^0)}}.
\label{eq:occupation_covariance_appendix}
\end{equation}
For Fourier amplitudes normalized consistently with the kinetic state, their equal-time equilibrium covariance is the identity. Linearized kinetic regression gives the two-time covariance $\langle\boldsymbol\varphi(q,t)\boldsymbol\varphi(-q,0)^{\mathsf T}\rangle=e^{-\boldsymbol{\mathcal K}(q)t}$ for $t\geq0$. This propagator describes equilibrium correlations, and stationary fluctuations are maintained by the corresponding stochastic forcing.

The identity $(\hbar\omega_\alpha)^2n_\alpha^0(1+n_\alpha^0)=k_{\rm B}T_0^2c_\alpha$ identifies the normalized energy coordinate as $\Theta=\mathbf e^{\mathsf T}\boldsymbol\varphi$. With the extensive current Fourier amplitude
\begin{equation}
J_i(q,t)=\int_Vd^3r\,e^{-i\mathbf q\cdot\mathbf r}j_i(\mathbf r,t),
\label{eq:fourier_current_appendix}
\end{equation}
the current is $J_i=\sqrt{Ck_{\rm B}T_0^2}\,\mathbf g_i^{\mathsf T}\boldsymbol\varphi$, with $\mathbf g_i=\mathbf Q\mathbf V_i|e\rangle$ the current vertex along direction $i$ ($\mathbf g_x=\mathbf g$). The temperature fluctuation is $\delta T(q)=\sqrt{k_{\rm B}T_0^2/C}\,\Theta$, consistent with the Fourier normalization in Eq.~\eqref{eq:KTT_main}. Its equilibrium autocorrelation is
\begin{subequations}
\label{eq:KTT_appendix}
\begin{align}
K_{TT}(q,t)&=\frac{k_{\rm B}T_0^2}{c_V}\mathbf e^{\mathsf T}e^{-\boldsymbol{\mathcal K}(q)t}\mathbf e,
\label{eq:KTT_time_appendix}\\
K_{TT}(q,0)&=\frac{k_{\rm B}T_0^2}{c_V}.
\label{eq:KTT_equal_time_appendix}
\end{align}
\end{subequations}
Therefore,
\begin{equation}
F_T(q,t)=\mathbf e^{\mathsf T}e^{-\boldsymbol{\mathcal K}(q)t}\mathbf e,
\qquad
\tilde F_T(q,s)=\mathbf e^{\mathsf T}[s\mathbf I+\boldsymbol{\mathcal K}(q)]^{-1}\mathbf e.
\label{eq:GT_matrix_methods_short}
\end{equation}
These expressions identify the normalized equilibrium temperature correlation as the energy-sector matrix element of the full kinetic propagator.

\subsection{Energy-conservation feedback and the Schur complement}
\label{app:current_feedback}

The energy row of Eq.~\eqref{eq:E_methods_appendix} is the continuity equation
\begin{equation}
c_V\partial_t\delta T+\boldsymbol\nabla\cdot\mathbf j=0.
\label{eq:energy_conservation_appendix}
\end{equation}
With $\mathbf A=s\mathbf I+\boldsymbol\Xi(q)$, the Schur complement of the full kinetic resolvent gives
\begin{equation}
\begin{aligned}
\tilde F_T(q,s)
&=\left[s-(iq)^2\mathbf g^{\mathsf T}\mathbf A^{-1}\mathbf g\right]^{-1}\\
&=\left[s+q^2\mathbf g^{\mathsf T}\mathbf A^{-1}\mathbf g\right]^{-1},
\end{aligned}
\end{equation}
which yields Eq.~\eqref{eq:temperature_green_main} on substituting Eq.~\eqref{eq:projected_kernel_derivation_appendix}.
For arbitrary propagation direction, $\mathbf q\cdot\overleftrightarrow{\tilde Z}\cdot\mathbf q$ replaces $q^2\tilde Z_L$. This identity connects the two spectral levels exactly within linearized kinetics: the projected resolvent generates the constitutive memory, while the full resolvent incorporates its feedback on the conserved energy coordinate.

The kinetic block of the full resolvent is
\begin{equation}
[(s\mathbf I+\boldsymbol{\mathcal K})^{-1}]_{QQ}=\mathbf A^{-1}-q^2\mathbf A^{-1}\mathbf g\tilde F_T\mathbf g^{\mathsf T}\mathbf A^{-1}.
\label{eq:full_QQ_resolvent_appendix}
\end{equation}
The ordinary equilibrium current correlation and projected constitutive memory are therefore
\begin{equation}
\begin{aligned}
\mathcal C_{J,ij}(q,t)&=c_V\mathbf g_i^{\mathsf T}e^{-\boldsymbol{\mathcal K}(q)t}\mathbf g_j,\\
Z_{ij}(q,t)&=c_V\mathbf g_i^{\mathsf T}e^{-\boldsymbol\Xi(q)t}\mathbf g_j,
\end{aligned}
\label{eq:ordinary_projected_correlations_appendix}
\end{equation}
where $\mathcal C_{J,ij}=\langle J_i(q,t)J_j(-q,0)\rangle/(Vk_{\rm B}T_0^2)$. Although the current vertices are orthogonal to the conserved energy mode, propagation under the full generator can pass through that coordinate. Contracting Eq.~\eqref{eq:full_QQ_resolvent_appendix} gives
\begin{equation}
\tilde{\mathcal C}_{J,L}(q,s)=s\tilde Z_L(q,s)\tilde F_T(q,s).
\label{eq:current_feedback_appendix}
\end{equation}
At $q=0$, $\tilde F_T=1/s$ and $\tilde{\mathcal C}_{J,L}=\tilde Z_L$, recovering the Green--Kubo relation. At finite $q$, the ordinary current correlation includes energy-conservation feedback and generally differs from the projected constitutive kernel. In particular, if $\tilde Z_L(q,0)$ is finite and nonzero, then $\tilde{\mathcal C}_{J,L}(q,s)\to0$ as $s\to0$, even though $\tilde Z_L(q,0)$ remains finite. Integrating the ordinary finite-$q$ current correlation cannot yield the static nonlocal conductivity.

This distinction also clarifies the approximation relating projected memory kernels to ordinary current correlations in Ref.~\cite{zeng_and_dong_2026_unified}. The replacement is exact at $q=0$, but at finite $q$ requires neglecting the energy-feedback contribution in Eq.~\eqref{eq:current_feedback_appendix}.

\subsection{Temperature poles and residues}

Writing $Q_n=q^2R_n/c_V$ in the channel form of Eq.~\eqref{eq:temperature_green_main}, the temperature poles satisfy
\begin{equation}
p_k+\sum_n\frac{Q_n}{p_k+a_n}=0, \label{eq:temperature_poles_appendix}
\end{equation}
after removable singularities are excluded. For simple temperature poles, $F_T(q,t)=\sum_kr_ke^{p_kt}$ [Eq.~\eqref{eq:temperature_pole_spectrum}], with normalized residues
\begin{equation}
r_k=\frac{1}{1-\sum_nQ_n/(p_k+a_n)^2}. \label{eq:KTT_residue_appendix}
\end{equation}
For $N$ distinct constitutive poles with nonzero residues and no cancellations, multiplying the secular equation by $\prod_n(p+a_n)$ gives a polynomial of degree $N+1$. The additional dynamical coordinate is the conserved energy mode excluded from $\boldsymbol\Xi$.

At $q=0$, energy conservation gives $\tilde F_T(0,s)=1/s$, corresponding to a stationary temperature fluctuation. For a regular constitutive kernel with finite bulk thermal conductivity $\kappa$, this conserved mode evolves into the diffusive temperature pole $p\simeq-D_0q^2$ of Sec.~\ref{sec:limiting}.

For numerical pole analysis, diagonalizing the projected generator first transforms the full generator into the complex-symmetric arrowhead form
\begin{equation}
\boldsymbol{\mathcal K}'(q)=
\begin{pmatrix}
0&iq\boldsymbol\beta^{\mathsf T}\\
iq\boldsymbol\beta&\operatorname{diag}(a_n)
\end{pmatrix},
\qquad
\beta_n^2=\frac{R_n}{c_V}. \label{eq:arrowhead_appendix}
\end{equation}
For nondefective eigenvectors $\mathbf y_k$, the temperature poles are $p_k=-\operatorname{eig}_k(\boldsymbol{\mathcal K}')$, and their residues are $r_k=y_{k,0}^2/(\mathbf y_k^{\mathsf T}\mathbf y_k)$. The complex square roots defining $\beta_n$ can be chosen consistently because the resolvent depends on $\beta_n^2$. This construction is equivalent to the secular equation and makes explicit the coupling of the conserved energy coordinate to each heat-current-active transport channel.

\subsection{Single-channel dynamics}

For a single relaxing constitutive kernel, $\tilde Z_L(q,s)=c_Vv_*^2/(s+\Gamma)$, the temperature correlation in the oscillatory regime is
\begin{equation}
F_T(q,t)=e^{-\Gamma t/2}\left[\cos(\omega_*t)+\frac{\Gamma}{2\omega_*}\sin(\omega_*t)\right], \label{eq:single_KTT_time_appendix}
\end{equation}
where $\omega_*=(q^2v_*^2-\Gamma^2/4)^{1/2}$.

\subsection{Real-frequency spectrum and Onsager regression}
\label{app:regression}

The normalized equilibrium temperature correlation satisfies $F_T(q,0)=1$. For an inversion-symmetric crystal, its real-frequency spectrum is even and nonnegative, with
\begin{equation}
\int_{-\infty}^{\infty}\operatorname{Re}\tilde F_T(q,i\omega)\,d\omega=\pi. \label{eq:temperature_frequency_sum_appendix}
\end{equation}
Changing variables to $\omega=qu$ gives the unit-area normalization of $S_T$ defined in Eq.~\eqref{eq:temperature_speed_spectrum}.
As for $S_Z$, the normalization holds distributionally when undamped poles generate delta-function contributions. The temperature representation is determined by the poles and residues of the full energy-sector resolvent, which differ from those of the projected transport generator.

The dynamic temperature-fluctuation spectrum $S_{TT}(q,\omega)$ of Sec.~\ref{sec:fluctuations} is the Fourier transform of $K_{TT}(q,t)$, and its relation to $\operatorname{Re}\tilde F_T$ uses the equilibrium symmetry of the correlation function. Thus, $S_T(q,u)$ is the positive-frequency, velocity-scaled representation of the normalized equilibrium fluctuation spectrum.

The regression relation~\eqref{eq:Onsager_regression_main} applies to a modulation prepared in local equilibrium. An ideal transient thermal grating prepares one Fourier component, whereas general bulk excitations superpose wave vectors and finite samples additionally require boundary conditions. Nonthermal initial phonon populations may introduce contributions beyond this temperature response.

For $q\ne0$, the constitutive kernel can be reconstructed formally from the complete temperature correlation:
\begin{equation}
\tilde Z_L(q,s)=\frac{c_V}{q^2}\left[\frac{1}{\tilde F_T(q,s)}-s\right]. \label{eq:inversion_appendix}
\end{equation}
This inversion determines the longitudinal constitutive response but not the full collision operator, heat-current-dark modes, or transverse transport. In practice, finite bandwidth and noise make spectral reconstruction ill-conditioned.

\section{Reduced transport descriptions and their validity}
\label{app:limiting_reductions}
\label{app:boundaries}

The following derivations identify the leading corrections to each reduced kernel and the resulting validity scale, the period at which the reduced kernel departs from the full kernel by a relative tolerance $\delta$. Each estimate is an expansion, built from the collision operator and the phonon velocities, in the part of the kernel that the reduction drops: long-wavelength expansions for Fourier and G--K, and expansions about RTA for RTA and collisionless streaming. The collision-sector matrices and streaming vertices use the notation of Appendix~\ref{app:methods_projected_transport}.

\subsection{Fourier diffusion}
\label{app:fourier}

When the slowest temperature channel is a real pole $p_{\rm slow}=-\Gamma_D$, which defines $D(q)=\Gamma_D/q^2$ (Sec.~\ref{sec:diagnostics}), expanding the kernel at this pole to first order in its rate gives
\begin{subequations}
\label{eq:slow_pole_appendix}
\begin{align}
\frac{D(q)}{D_0}&\simeq\frac{K(q)}{1-\epsilon_{\rm mem}(q)}, \label{eq:slow_pole_ratio_appendix}\\
K&=\frac{\tilde Z_L(q,0)}{\kappa},\qquad \epsilon_{\rm mem}=q^2\sum_n\frac{R_n}{c_Va_n^2}, \label{eq:slow_pole_terms_appendix}
\end{align}
\end{subequations}
with static factor $K$ and memory correction $\epsilon_{\rm mem}$. Eliminating the even sector of the projected generator (Appendix~\ref{app:projected}) gives the static kernel
\begin{equation}
\tilde Z_L(q,0)=c_V\boldsymbol\Lambda_x^{\mathsf T}\left[\boldsymbol\gamma+q^2\boldsymbol\Pi_x\boldsymbol\varepsilon^{-1}\boldsymbol\Pi_x^{\mathsf T}\right]^{-1}\boldsymbol\Lambda_x ,
\label{eq:static_schur_appendix}
\end{equation}
and because $q^2\boldsymbol\Pi_x\boldsymbol\varepsilon^{-1}\boldsymbol\Pi_x^{\mathsf T}$ is positive semidefinite, $K\le1$ at every $q$. The static factor lowers the slow decay rate and a positive memory correction raises it, so within this approximation $D(q)>D_0$ when $\epsilon_{\rm mem}>1-K$. To leading order $1-K=(q\ell_s)^2$ and $\epsilon_{\rm mem}=(q\ell_m)^2$, with
\begin{subequations}
\label{eq:Fourier_lengths_appendix}
\begin{align}
\ell_m^2&=\boldsymbol\Lambda_x^{\mathsf T}\boldsymbol\gamma^{-2}\boldsymbol\Lambda_x, \label{eq:memory_length_appendix}\\
\ell_s^2&=\frac{1}{D_0}\boldsymbol\Lambda_x^{\mathsf T}\boldsymbol\gamma^{-1}\boldsymbol\Pi_x\boldsymbol\varepsilon^{-1}\boldsymbol\Pi_x^{\mathsf T}\boldsymbol\gamma^{-1}\boldsymbol\Lambda_x, \label{eq:static_length_appendix}
\end{align}
\end{subequations}
the memory and static-nonlocality lengths of the kernel, so $D(q)/D_0\simeq1+q^2(\ell_m^2-\ell_s^2)$. At this real pole the two corrections subtract and can cancel, so the slow decay rate can match Fourier while the kernel does not. The validity test compares the kernels on the imaginary axis, over the frequency band that the temperature response samples,
\begin{equation}
\frac{\tilde Z_L(q,i\omega)}{\kappa}\simeq1-(q\ell_s)^2-i\omega\frac{\ell_m^2}{D_0}, \label{eq:Z_expansion_appendix}
\end{equation}
where the static correction is real and the memory correction imaginary, so they add in quadrature. The memory correction grows with $\omega$. The band ends at the median frequency $\omega_T$ of the temperature power spectrum, which for a diffusive response is $\omega_T=D_0q^2$, where the memory correction reaches $(q\ell_m)^2$. The largest error on the band is $q^2(\ell_s^4+\ell_m^4)^{1/2}$, and Fourier transport holds within $\delta$ for
\begin{equation}
L>L_F=2\pi\,\frac{(\ell_s^4+\ell_m^4)^{1/4}}{\sqrt\delta}. \label{eq:LF_appendix}
\end{equation}

\subsection{Guyer--Krumhansl reduction}
\label{app:GK}
\label{sec:methods_GK_reference}

At $q=0$, the symmetrized collision operator is Hermitian, with orthonormal odd-parity eigenmodes $|\chi_\lambda\rangle$. Their momentum character is quantified by the overlap with the three-dimensional crystal-momentum subspace,
\begin{subequations}
\label{eq:momentum_character_appendix}
\begin{align}
H_\lambda&=\sum_{i=x,y,z}|\langle d_i|\chi_\lambda\rangle|^2,\\
d_{\alpha i}&\propto k_{\alpha i}\sqrt{n_\alpha^0(1+n_\alpha^0)},
\end{align}
\end{subequations}
where $\alpha$ labels phonon states and the momentum vectors $|d_i\rangle$ are orthonormalized. The overlap satisfies $0\leq H_\lambda\leq1$, with $H_\lambda\simeq1$ indicating that the collision eigenmode lies predominantly within the crystal-momentum subspace. For a degenerate eigenspace $\mathcal E$, the basis-independent total momentum overlap is $\mathrm{Tr}(\boldsymbol{\mathcal P}_M\boldsymbol{\mathcal P}_{\mathcal E})$, where $\boldsymbol{\mathcal P}_M=\sum_i|d_i\rangle\langle d_i|$ and $\boldsymbol{\mathcal P}_{\mathcal E}$ is the projector onto the eigenspace.

The phonon-space participation ratio is defined as
\begin{equation}
P_\lambda=\frac{\left(\sum_\alpha|\chi_{\lambda\alpha}|^2\right)^2}{\sum_\alpha|\chi_{\lambda\alpha}|^4}.
\label{eq:participation_ratio_appendix}
\end{equation}
It measures the effective number of phonon states participating in a collision eigenmode, ranging from unity for a mode localized on one phonon state to the total number of phonon states for a uniformly distributed mode. A large participation ratio indicates broad phonon participation, whereas a large momentum overlap identifies alignment with crystal momentum. Within degenerate multiplets, individual participation ratios are evaluated in the momentum-aligned basis. Heat-current dominance and spectral separation are assessed separately below.

At each temperature, we select the odd collision mode with the largest momentum overlap in the longitudinal symmetry sector and denote it by $h$, with $\Gamma_h=\gamma_h$. For $\mathbf q=q\hat x$, define the longitudinal current coupling $\Lambda_h=\hat{\mathbf q}\cdot\boldsymbol\Lambda_h$. The vectors $\boldsymbol\Lambda_\lambda$ and $\boldsymbol\Pi_{\lambda\iota}$ are defined in Appendix~\ref{app:methods_projected_transport}, and $\varepsilon_\iota$ denotes the relaxation rate of an even collision mode orthogonal to energy. The selected mode contributes
\begin{equation}
\kappa_h=c_V\frac{|\Lambda_h|^2}{\Gamma_h}
\label{eq:kappa_hydrodynamic_appendix}
\end{equation}
to the bulk conductivity. A single-mode description of bulk heat conduction requires $\kappa_h/\kappa\simeq1$.

When $h$ is spectrally isolated and nondegenerate within this symmetry sector, streaming couples it perturbatively to the even collision modes. The couplings $|q\hat{\mathbf q}\cdot\boldsymbol\Pi_{h\iota}|$ must be small relative to the gaps $|\varepsilon_\iota-\Gamma_h|$, with a small total admixture of the eliminated modes. Second-order perturbation theory for the energy-projected kinetic generator gives
\begin{equation}
a_h(q)=\Gamma_h+\mu q^2+O(q^4),
\label{eq:GK_eigenvalue_appendix}
\end{equation}
where
\begin{equation}
\mu=\sum_\iota\frac{|\hat{\mathbf q}\cdot\boldsymbol\Pi_{h\iota}|^2}{\varepsilon_\iota-\Gamma_h}.
\label{eq:mu_methods_short}
\end{equation}
The conserved energy mode is excluded from this sum because it is treated separately through energy conservation. The correction is positive when the coupled even-sector rates exceed $\Gamma_h$.

For comparison with the full calculation, we construct a conductivity-normalized G--K reference using Eq.~\eqref{eq:GK_kernel_main}. The relaxation rate $\Gamma_h$ and nonlocal coefficient $\mu$ are determined from the $q=0$ collision spectrum without fitting finite-$q$ transport modes, while the residue is set to $\kappa\Gamma_h$, with $\kappa$ the full bulk conductivity. This normalization preserves the exact long-wavelength diffusivity $D_0$. The microscopic hydrodynamic channel instead has the residue $c_V|\Lambda_h|^2=\kappa_h\Gamma_h$. Thus, the reference residue and characteristic speed exceed their microscopic counterparts by factors $\kappa/\kappa_h$ and $(\kappa/\kappa_h)^{1/2}$, respectively. The conductivity-normalized reference approaches the microscopic single-channel description when $\kappa_h/\kappa\simeq1$, although finite-$q$ changes in the microscopic channel residue are not retained. The corresponding longitudinal G--K equation is $(1+\tau\partial_t-\ell^2\nabla^2)j_L=-\kappa\partial_xT$.

Normalized to its own instantaneous weight $\kappa\Gamma_h$, the G--K constitutive spectrum is a zero-centered Lorentzian of half-width $\gamma_q/q$,
\begin{equation}
S_Z^{\rm GK}(q,u)=\frac{2}{\pi}\frac{q\gamma_q}{\gamma_q^2+q^2u^2},
\label{eq:GK_SZ_main}
\end{equation}
and the temperature-fluctuation spectrum of Eq.~\eqref{eq:GK_FT_main} is
\begin{equation}
S_T^{\rm GK}(q,u)=\frac{2}{\pi}\frac{\gamma_q v_h^2/q}{(v_h^2-u^2)^2+(\gamma_q u/q)^2}.
\label{eq:GK_ST_main}
\end{equation}
Unlike the constitutive Lorentzian, $S_T^{\rm GK}$ develops a finite-frequency maximum when $\gamma_q<\sqrt{2}\,qv_h$, which is stricter than the pole condition of Eq.~\eqref{eq:GK_second_sound_condition} and possible only if $\mu<D_0/2$. The normalization of Eq.~\eqref{eq:GK_SZ_main} differs from the microscopic convention in Eq.~\eqref{eq:constitutive_speed_spectrum} whenever $\kappa\Gamma_h\ne c_Vv_{\rm rms}^2$. In the microscopic normalization, the G--K spectrum is multiplied by $\kappa\Gamma_h/(c_Vv_{\rm rms}^2)$, the ratio of the reference's instantaneous constitutive weight to that of the full microscopic kernel.

The temperature poles of Eq.~\eqref{eq:GK_FT_main} are the roots of $p^2+\gamma_qp+D_0\Gamma_hq^2=0$ and form a damped second-sound pair when $\gamma_q<2qv_h$. For $0<\mu<D_0$ this holds between
\begin{equation}
q_\pm=\frac{\sqrt{D_0\Gamma_h}\pm\sqrt{\Gamma_h(D_0-\mu)}}{\mu}.
\label{eq:GK_q_bounds}
\end{equation}
The stronger condition $|\operatorname{Im}p|>|\operatorname{Re}p|$ is equivalent here to the finite-frequency-maximum condition $\gamma_q<\sqrt2\,qv_h$ above.

\emph{Validity.} Where the momentum-organized sector is spectrally isolated and carries nearly all the bulk conductivity, the conductivity-normalized G--K reference reproduces the leading temporal memory of the full kernel, with $\ell_m^2\simeq D_0/\Gamma_h$. Its static nonlocality generally differs from the microscopic result: the G--K static factor $1/(1+q^2\ell^2)\simeq1-q^2\ell^2$, with $\ell^2=\mu/\Gamma_h$, replaces $1-(q\ell_s)^2$. The leading relative difference between the static kernels is therefore $q^2|\ell^2-\ell_s^2|$. Requiring this difference to remain below $\delta$ gives
\begin{equation}
L>L_{\rm GK}=2\pi\sqrt{\frac{|\ell^2-\ell_s^2|}{\delta}},
\label{eq:LGK_appendix}
\end{equation}
within the small-$q$ regime where the expansion applies. This static-kernel accuracy scale is distinct from the oscillation window $q_\pm$, which characterizes the reduced model rather than its microscopic validity.

\subsection{Energy-conserving relaxation-time approximation}
\label{sec:methods_ecRTA}

A diagonal relaxation operator $\boldsymbol\Gamma_{\rm ph}=\operatorname{diag}(\gamma_\alpha)$, with $\gamma_\alpha=\tau_\alpha^{-1}$, does not conserve energy. Relaxation toward a local equilibrium constrained by energy conservation instead gives
\begin{equation}
\boldsymbol\Omega_{\rm RTA} = \boldsymbol\Gamma_{\rm ph}-\frac{\boldsymbol\Gamma_{\rm ph}|e\rangle\langle e|\boldsymbol\Gamma_{\rm ph}}{\langle e|\boldsymbol\Gamma_{\rm ph}|e\rangle}.
\label{eq:ecRTA_operator_appendix}
\end{equation}
This rank-one-corrected operator is positive semidefinite and satisfies $\boldsymbol\Omega_{\rm RTA}|e\rangle=0$. Throughout this work RTA denotes this energy-conserving form. Because $|e\rangle$ and $\boldsymbol\Gamma_{\rm ph}|e\rangle$ are even under $x\to-x$, the correction leaves the odd-parity block equal to $\boldsymbol\Gamma_{\rm ph}$, so the bulk conductivity $\kappa_{\rm RTA}$ is the standard RTA value, but it changes the even-parity dynamics that couple to the heat current at finite $q$. The corresponding projected kinetic generator and constitutive kernel are
\begin{equation}
\begin{aligned}
\boldsymbol\Xi_{\rm RTA}(q) &=\left.\mathbf Q[\boldsymbol\Omega_{\rm RTA}+iq\mathbf V_L]\mathbf Q\right|_{\operatorname{Ran}\mathbf Q},\\
\tilde Z_L^{\rm RTA}(q,s) &=c_V\mathbf g^{\mathsf T}[s\mathbf I+\boldsymbol\Xi_{\rm RTA}(q)]^{-1}\mathbf g.
\end{aligned}
\label{eq:ecRTA_projected_appendix}
\end{equation}
The normalized equilibrium temperature correlation follows from the full conserving generator, or equivalently from the exact energy-conservation relation $\tilde F_T^{\rm RTA}(q,s)=1/[s+q^2\tilde Z_L^{\rm RTA}(q,s)/c_V]$.

An explicit analytical representation follows from the rank-one structure of the conserving correction. Define the heat-capacity weights $\hat c_\alpha=c_\alpha/C$, the diagonal phonon resolvents $h_\alpha(q,s)=1/(s+\gamma_\alpha+iqv_{\alpha L})$, and the weighted sums
\begin{equation}
\begin{aligned}
A(q,s)&=\sum_\alpha \hat c_\alpha h_\alpha(q,s),\\
B(q,s)&=\sum_\alpha \hat c_\alpha\gamma_\alpha h_\alpha(q,s),\\
C_2(q,s)&=\sum_\alpha \hat c_\alpha\gamma_\alpha^2h_\alpha(q,s),\\
\bar\gamma&=\sum_\alpha \hat c_\alpha\gamma_\alpha.
\end{aligned}
\label{eq:ecRTA_auxiliary_appendix}
\end{equation}
Applying the Sherman--Morrison identity to the full kinetic resolvent, with $\mathbf M=s\mathbf I+\boldsymbol\Gamma_{\rm ph}+iq\mathbf V_L$, gives
\begin{equation}
[s\mathbf I+\boldsymbol{\mathcal K}_{\rm RTA}(q)]^{-1} = \mathbf M^{-1} + \frac{\mathbf M^{-1}\boldsymbol\Gamma_{\rm ph}|e\rangle \langle e|\boldsymbol\Gamma_{\rm ph}\mathbf M^{-1}} {\bar\gamma-\langle e|\boldsymbol\Gamma_{\rm ph}\mathbf M^{-1}\boldsymbol\Gamma_{\rm ph}|e\rangle}. \label{eq:ecRTA_SM_appendix}
\end{equation}
Projection onto the conserved energy coordinate yields
\begin{equation}
\tilde F_T^{\rm RTA}(q,s) = A(q,s)+\frac{B(q,s)^2}{\bar\gamma-C_2(q,s)}. \label{eq:ecRTA_FT_analytic_appendix}
\end{equation}
At $q=0$, the identities $B=1-sA$ and $\bar\gamma-C_2=sB$ give $\tilde F_T^{\rm RTA}(0,s)=1/s$, confirming exact energy conservation. The constitutive kernel follows from the inversion relation, Eq.~\eqref{eq:inversion_appendix}, with $\tilde F_T^{\rm RTA}$.
These expressions provide analytical RTA references for both spectral levels directly from phonon velocities, heat capacities, and relaxation rates, without numerical diagonalization. They retain the distribution of phonon streaming frequencies and relaxation rates while neglecting the microscopic off-diagonal collision couplings of the full scattering operator.

The normalized spectra $S_Z^{\rm RTA}$ and $S_T^{\rm RTA}$ follow from Eqs.~\eqref{eq:constitutive_speed_spectrum} and \eqref{eq:temperature_speed_spectrum} with these kernels. Both have unit integrated spectral weight, but neither generally reduces to a single Lorentzian. Although the diagonal operator $\boldsymbol\Gamma_{\rm ph}+iq\mathbf V_L$ has poles at $s=-\gamma_\alpha-iqv_{\alpha L}$, these are not generally the poles of the projected constitutive kernel: energy projection and the conserving correction, which couples phonon populations through the energy coordinate, reorganize the response, so projected RTA channels need not be phonon diagonal or possess positive individual residues.

\emph{Validity.} We use the normalized kernel $z=Z_L(q,t)/Z_L(q,0)$ and scaled time $\theta=qv_{\rm rms}t$. The window norm $\|\cdot\|_w$ and distance $d_{Z,w}^X$ are defined in Sec.~S3\,A. RTA has no long-wavelength validity boundary, because $\kappa_{\rm RTA}\ne\kappa$. Near free streaming, it departs from the full theory only through the in-scattering it omits, $\boldsymbol\Omega_{\rm od}=\boldsymbol\Omega-\boldsymbol\Omega_{\rm RTA}$. Let $\mathbf R_0=[s\mathbf I+\boldsymbol\Omega_{\rm RTA}+iq\mathbf V_L]^{-1}$, $\mathbf y=\mathbf R_0|e\rangle$ and $F_0=F_T^{\rm RTA}$. The Dyson expansion $F_T=F_0+F_1+F_2+\cdots$ has the Laplace-domain terms
\begin{equation}
\begin{aligned}
\tilde F_0 &=\langle e|\mathbf R_0|e\rangle,\\
\tilde F_1 &=-\mathbf y^{\mathsf T}\boldsymbol\Omega_{\rm od}\mathbf y,\\
\tilde F_2 &=\mathbf y^{\mathsf T}\boldsymbol\Omega_{\rm od}\mathbf R_0\boldsymbol\Omega_{\rm od}\mathbf y.
\end{aligned}
\label{eq:rta_dyson}
\end{equation}
Expanding the inversion relation~\eqref{eq:inversion_appendix} gives $z=z_{\rm RTA}+z_1^{\rm od}+z_2^{\rm od}+\cdots$, where
\begin{equation}
\begin{aligned}
\tilde z_1^{\rm od} &=-\frac{1}{q^2v_{\rm rms}^2}\frac{\tilde F_1}{\tilde F_0^2},\\
\tilde z_2^{\rm od} &=\frac{1}{q^2v_{\rm rms}^2}\left(\frac{\tilde F_1^2}{\tilde F_0^3}-\frac{\tilde F_2}{\tilde F_0^2}\right).
\end{aligned}
\label{eq:rta_kernel_corrections}
\end{equation}
Here the Laplace transforms are with respect to physical time $t$, and $Z_L(q,0^+)=c_Vv_{\rm rms}^2$. The quadratic term includes both the second-order change of $F_T$ and the quadratic contribution from its first-order change.

Over the first free-flight oscillation, $t=\theta L/(2\pi v_{\rm rms})$, the first-order distance grows in proportion to the period, $d_{Z,w}^{\rm RTA}\simeq\gamma_{\rm od}L/(2\pi v_{\rm rms})$. This defines $\gamma_{\rm od}$, the in-scattering rate of the current weighted by its dephasing over the window, whose scale is the difference between the out-scattering rate of the current, $\bar\gamma_J=\sum_\alpha c_\alpha v_{\alpha L}^2\gamma_\alpha/\sum_\alpha c_\alpha v_{\alpha L}^2$, and its net decay rate under the full operator, $\bar\gamma_J^{\rm full}=\boldsymbol\Lambda_x^{\mathsf T}\boldsymbol\gamma\boldsymbol\Lambda_x/\boldsymbol\Lambda_x^{\mathsf T}\boldsymbol\Lambda_x$. The RTA holds for
\begin{equation}
L<L_{\rm RTA}\simeq\frac{2\pi\delta\,v_{\rm rms}}{\gamma_{\rm od}}.
\label{eq:LRTA_appendix}
\end{equation}
Where collisions return most of the current, this first-order value overestimates $L_{\rm RTA}$, and the boundary is evaluated with $z_2^{\rm od}$ included.

\subsection{Collisionless streaming}
\label{sec:methods_collisionless}

At a fixed temperature, decreasing $L$ increases the phonon streaming frequencies relative to the collision rates. When scattering becomes negligible over the characteristic evolution time of the temperature modulation, the full kinetic generator is dominated by streaming, $\boldsymbol{\mathcal K}(q)\simeq iq\mathbf V_L$, as in Sec.~\ref{sec:limiting}.
The collision operator remains finite, but its influence on the relevant phonon populations becomes small over the period $L$, so the full kinetic theory and RTA both approach free streaming, though not at the same rate: the full theory returns part of the scattered current, which RTA removes (below). The normalized temperature correlation then approaches $F_T^{\rm cl}$ of Eq.~\eqref{eq:collisionless_FT_main}, and the constitutive kernel follows from the inversion relation in Eq.~\eqref{eq:inversion_appendix}. For a discrete velocity set, the kernel poles lie at the zeros of $\tilde F_T^{\rm cl}$, with nonzero frequencies interlacing the distinct positive phonon streaming frequencies $q|v_{\alpha L}|$; a pole at zero frequency can also occur, as in the two-velocity example of Sec.~\ref{sec:limiting}. The resulting projected channels are undamped, but their frequencies and residues differ from the phonon streaming frequencies and the $c_\alpha v_{\alpha L}^2$ weights of the ordinary current correlation, while the instantaneous weight still satisfies $\sum_nR_n=c_V\langle e|\mathbf V_L^2|e\rangle$ [Eq.~\eqref{eq:constitutive_sum_rule}]. Only the initial modulation is in local equilibrium; the populations then stream far from it, and the exact constitutive kernel describes this evolution.

\emph{Validity.} The initial collisional decay of the current gives the leading departure of the full kernel from the collisionless kernel, at first order in time,
\begin{equation}
z(\theta)-z_{\rm cl}(\theta)=-\frac{\bar\gamma_J^{\rm full}}{qv_{\rm rms}}\,\theta+O(\theta^2).
\label{eq:cl_short_time}
\end{equation}
Where collisions return most of the scattered current, the net rate $\bar\gamma_J^{\rm full}$ is much smaller than the out-scattering rate $\bar\gamma_J$. An expansion in $\boldsymbol\Omega$ around free streaming does not converge at the edge at low temperature: the out-scattering accumulated over one free flight, about $\bar\gamma_Jt$, is much larger than the net change of the current, which the returned current reduces to $\delta$. The expansion~\eqref{eq:rta_dyson}, which sums the diagonal rates exactly, converges, and collisionless streaming holds for $L<L_{\rm cl}$, with
\begin{equation}
\|z_{\rm RTA}+z_1^{\rm od}+z_2^{\rm od}-z_{\rm cl}\|_w=\delta\,\|z_{\rm cl}\|_w\quad\text{at }L=L_{\rm cl}.
\label{eq:Lcl_appendix}
\end{equation}
The result lies between the out-scattering and net-decay estimates, $2\pi\delta v_{\rm rms}/\bar\gamma_J<L_{\rm cl}<2\pi\delta v_{\rm rms}/\bar\gamma_J^{\rm full}$. Where collisions return most of the current, $L_{\rm cl}>L_{\rm RTA}$: the full kernel is closer to free streaming than to RTA.

\emph{Temperature level.} The temperature response departs from free streaming more weakly than the kernel. With Laplace transforms taken in physical time, energy conservation gives $\tilde F_T=[s+q^2v_{\rm rms}^2\tilde z]^{-1}$, so a kernel change produces
\begin{equation}
\delta\tilde F_T=-q^2v_{\rm rms}^2\bigl[\tilde F_T^{\rm cl}(q,s)\bigr]^2\,\delta\tilde z,
\label{eq:cl_filter}
\end{equation}
which with Eq.~\eqref{eq:cl_short_time} gives $\delta F_T=(\bar\gamma_J^{\rm full}/qv_{\rm rms})\,\theta^3/6$ at short times: collisions reach the temperature only through the heat current.

\bibliography{unified_theory,PR_SM}

\end{document}


\title{Supplemental Material for ``Hidden Heat-Current Transport Channels Behind Observable Temperature Dynamics''}
\author{Yi Zeng}
\email{yi.zeng@nlr.gov}
\affiliation{National Laboratory of the Rockies, Golden, Colorado 80401, USA}
\author{Jianjun Dong}
\email{dongjia@auburn.edu}
\affiliation{Department of Physics, Auburn University, Auburn, Alabama 36830, USA}
\date{\today}
\maketitle
\raggedbottom

This Supplemental Material provides additional details on the first-principles calculations, numerical validation, and determination of transport-regime boundaries presented in the main text. Section~\ref{sec:S_methods} examines numerical convergence and compares the first-principles results for NaF with experimental measurements. Section~\ref{sec:S_reorg} supplements the analysis in Sec.~III of the main text, covering the momentum-organized sector and its limits (Sec.~\ref{sec:S_hydro}), the evolution toward distributed transport channels at 17~K (Sec.~\ref{sec:S_17K_evolution}), and relaxation and streaming beyond momentum organization at 25, 30, and 300~K (Sec.~\ref{sec:S_collective}). Finally, Sec.~\ref{sec:S_crossmap} defines the numerical criteria used to classify the constitutive spectra and identify the transport-regime boundaries shown in Fig.~6 of the main text.

\section{Numerical methods and validation}
\label{sec:S_methods}

\subsection{Mesh convergence}
\label{sec:S_checks}

\label{sec:S_numerics}
Unless stated otherwise, calculations are performed for intrinsic NaF on a $17\times17\times17$ phonon mesh, with $\mathbf q\parallel[100]$. Constitutive kernels are evaluated using the full projected resolvent, retaining contributions from all transport channels; weak channels are omitted only from graphical representations. Time-domain temperature correlation functions are evaluated from pole sums and independently verified by numerical Laplace inversion, with agreement better than $10^{-3}$. RTA and collisionless correlation functions are evaluated using the same inversion procedure. The streaming crossover length $L_s$ is located by bisection on a logarithmic length scale. Eigenmodes are tracked between neighboring grating periods using normalized Hermitian overlaps in a common kinetic basis, with nearly degenerate modes and complex-conjugate pairs compared as subspaces (main-text Appendix~A).

\begin{figure}[htbp]
\centering
\includegraphics[width=0.8\textwidth]{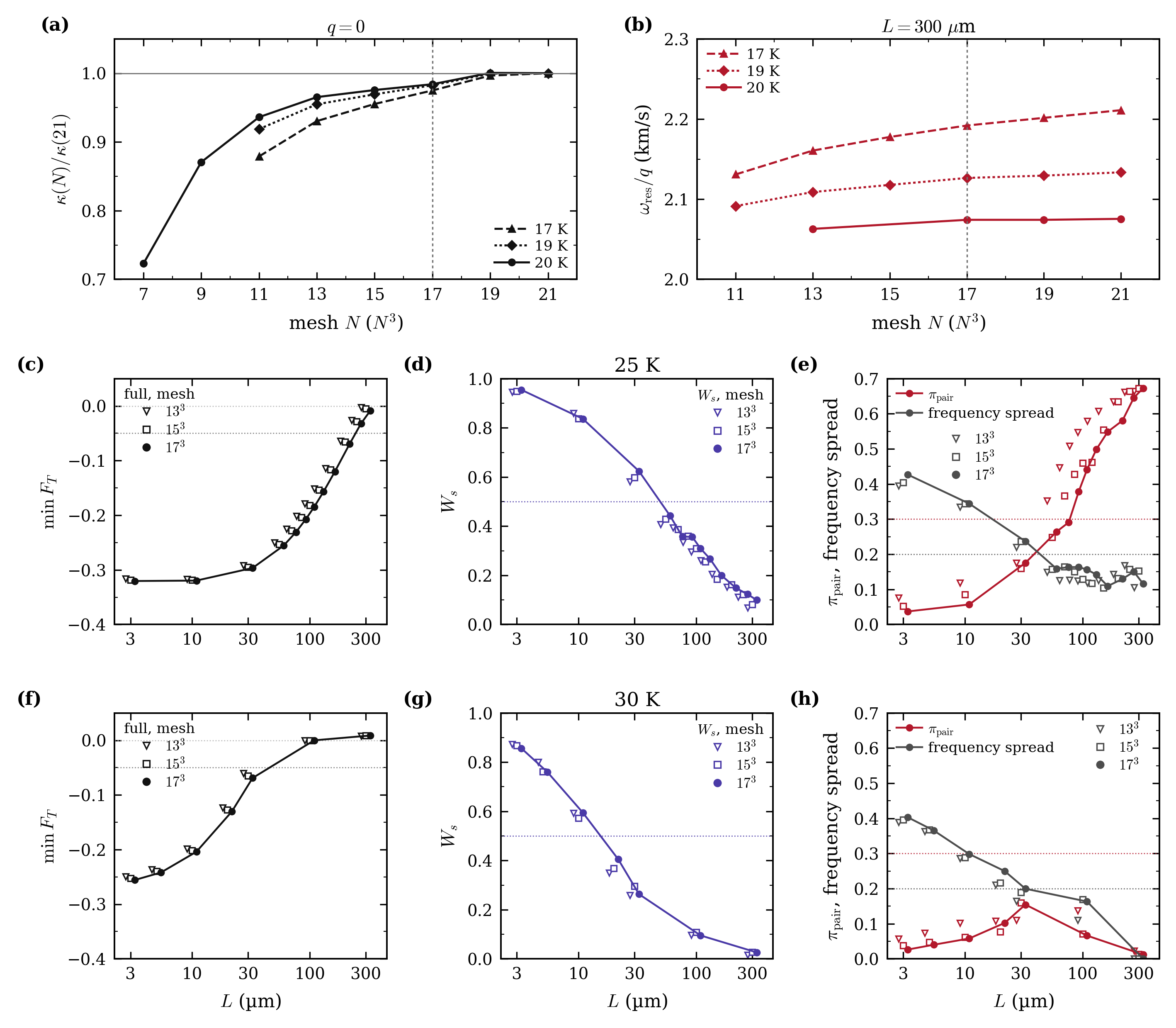}
\caption{Mesh checks. (a) Bulk conductivity of the full collision operator versus mesh $N^3$ at 17, 19 and 20~K, normalized to $N=21$; the $21^3$ values are $\kappa=93{,}400$, $30{,}000$ and $18{,}960$~W\,m$^{-1}$K$^{-1}$ at 17, 19 and 20~K. (b) Second-sound resonance speed $\omega_{\rm res}/q$ from the peak of $\operatorname{Re}\tilde F_T$ at $L=300~\mu$m at 17, 19 and 20~K. In (a,b) the vertical dotted line marks the $17^3$ mesh used throughout. (c)--(e) 25~K and (f)--(h) 30~K at $L=3$--$300~\mu$m on $13^3$ (triangles), $15^3$ (squares) and $17^3$ (filled circles, joined), offset horizontally for visibility: (c,f) minimum of the normalized temperature response; (d,g) streaming index $W_s$ (main-text Sec.~II\,D); (e,h) weight $\pi_{\rm pair}$ of the leading temperature-pole pair and relative frequency spread of the underdamped temperature poles (Sec.~\ref{sec:S_screen}). Dotted lines mark the thresholds.}
\label{fig:S_mesh}
\end{figure}

\label{sec:S_mesh}
On the $17^3$ mesh the bulk conductivity lies within 2.5\% of its $21^3$ value at 17--20~K [Fig.~\ref{fig:S_mesh}(a)]. At 25 and 30~K the response minima and $W_s$ on the $13^3$, $15^3$ and $17^3$ meshes agree within 0.01 and 0.06 at every sampled period. Pole shares converge more slowly: at 25~K the leading-pair share $\pi_{\rm pair}$ (defined in Sec.~\ref{sec:S_screen}) is up to 0.16 larger on $13^3$ than on $17^3$, so the pair passes the weight test from $55~\mu$m on $13^3$ and from $85~\mu$m on $17^3$, while from $55~\mu$m to $300~\mu$m the frequency spread stays at or below 0.17 on every mesh. The 25-K wave is present on all three meshes. The second-sound resonance speed at $300~\mu$m, about 2.2~km/s, changes by less than 0.1~km/s up to $21^3$ (3.8\% at 17~K and 2.0\% at 19~K from $11^3$, 0.6\% at 20~K from $13^3$), and the $17^3$ value lies within 0.9\% of the $21^3$ value [Fig.~\ref{fig:S_mesh}(b)]. For heat-pulse calculations at 14, 16, and 17~K, peak arrival speeds differ by less than 1\% between the $13^3$ and $17^3$ meshes for both slab thicknesses.

\label{sec:S_sumrule}
We validate the numerical calculations against three analytical sum rules derived in main-text Appendixes~A and B. The constitutive residue sum $\sum_n R_n=Z_L(q,0^+)$ remains independent of $q$ to within $4.6\times10^{-9}$. The constitutive spectrum $S_Z$ is nonnegative and satisfies its unit-area normalization to within $2.5\times10^{-4}$. The temperature-pole residues sum to unity to within $10^{-8}$.

\subsection{Experimental benchmarks}
\label{sec:S_experiments}

\begin{figure}[htbp]
\centering
\includegraphics[width=\textwidth]{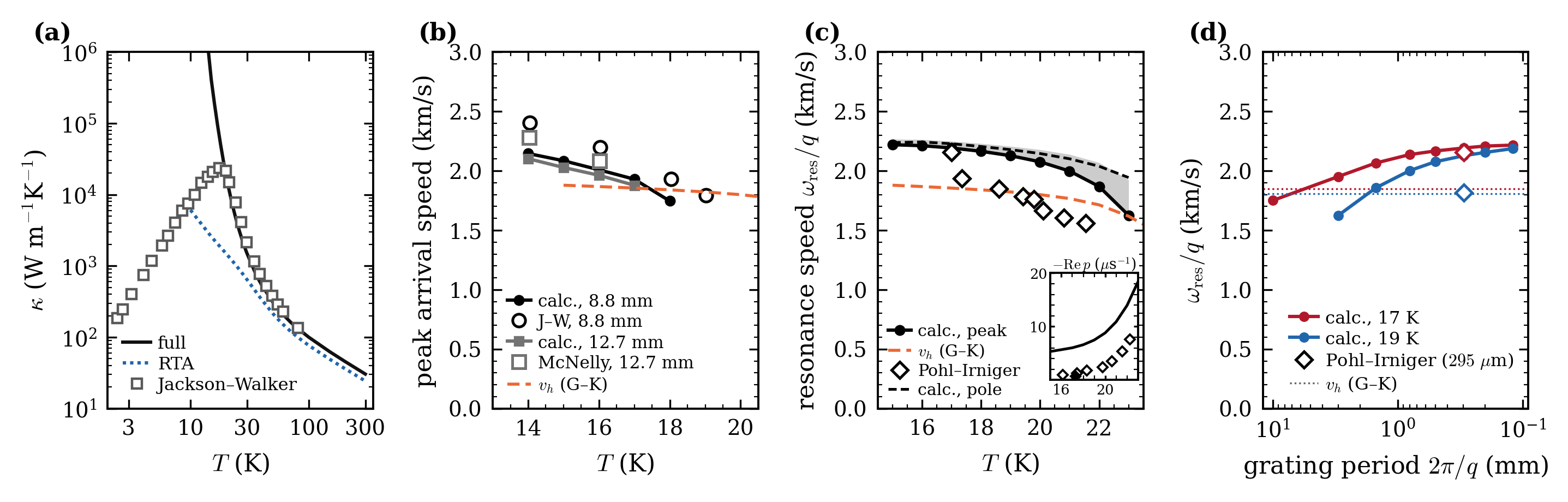}
\caption{\textbf{Experimental benchmarks for NaF.}
(a) Bulk thermal conductivity calculated with the full collision operator and RTA, compared with measurements on the purest crystal of Jackson and Walker~\cite{jackson1971thermal}.
(b) Heat-pulse peak arrival speeds calculated for propagation distances of 8.8 and 12.7~mm (solid lines), compared with measurements by Jackson and Walker~\cite{jackson1971thermal} (J--W, 8.8~mm) and McNelly \emph{et al.}~\cite{mcnelly1970heat} (12.7~mm).
(c) Light-scattering resonance speeds at a grating period of 295~$\mu$m. The solid curve gives the peak of $\operatorname{Re}\tilde F_T$, with shading extending to the peak of $|\tilde F_T|$ to indicate the dependence on detection phase. The black dashed curve gives the phase speed $|\operatorname{Im}p|/q$ of the leading temperature pole. Open diamonds are measured resonance speeds~\cite{pohl1976observation}. Inset: calculated leading-pole damping $-\operatorname{Re}p$ (solid curve), measured damping inferred from Eq.~(3) of Ref.~\cite{pohl1976observation} (open diamonds), and damping from a single-pole fit to the measured 17.3-K resonance (filled diamond). 
(d) Calculated resonance phase speed versus grating period at 17 and 19~K, compared with the measured value at 295~$\mu$m. The G--K hydrodynamic speed $v_h$ is shown dashed (orange) in (b,c) and dotted in (d). Conductivity data are digitized from Fig.~2 of Jackson and Walker~\cite{jackson1971thermal}, and heat-pulse and light-scattering data from Figs.~1(b), 3, and 4 of Pohl and Irniger~\cite{pohl1976observation}.}
\label{fig:S_experiments}
\end{figure}

We benchmark the first-principles phonon model against available measurements of bulk thermal conductivity, heat-pulse propagation, and light-scattering resonances in NaF (Fig.~\ref{fig:S_experiments}). These observables provide complementary tests of the calculated phonon properties, scattering rates, and finite-wave-vector transport dynamics. The calculations reproduce the principal trends in conductivity and second-sound propagation, including a short-wavelength thermal resonance beyond the G--K oscillatory window, but predict substantially broader light-scattering resonances than observed experimentally.

\emph{Bulk thermal conductivity.}
\label{sec:S_kappa}
Figure~\ref{fig:S_experiments}(a) compares the calculated bulk conductivity with measurements on the purest NaF crystal of Jackson and Walker~\cite{jackson1971thermal}. Below approximately 20~K, the measured conductivity is limited by impurity, dislocation, and boundary scattering, reaching a maximum of $24\,000$~W\,m$^{-1}$K$^{-1}$ near 17~K, whereas the calculated intrinsic bulk conductivity continues to increase upon cooling. Above 20~K, the calculation underestimates the measured conductivity by up to 40\% at 24--26~K, with the discrepancy decreasing to 3\% at 80~K. This difference indicates quantitative uncertainty in the calculated temperature dependence of intrinsic phonon scattering. The calculated longitudinal and transverse first-sound velocities along [100], 5.9 and 3.0~km/s, are also 4--6\% below the measured values of 6.17 and 3.19~km/s~\cite{hardy1971velocity}. These comparisons provide an indication of the accuracy of the first-principles phonon model; no empirical temperature rescaling is applied.

\emph{Second-sound propagation.}
\label{sec:S_second_sound}
Figures~\ref{fig:S_experiments}(b)--(d) benchmark the calculated second-sound propagation against heat-pulse and light-scattering measurements compiled by Pohl and Irniger~\cite{pohl1976observation}. Because these techniques probe different spatial and temporal scales, we calculate the corresponding experimental observables rather than comparing them with a single characteristic velocity.

For heat-pulse measurements~\cite{mcnelly1970heat,jackson1971thermal}, we calculate the local temperature at the far face of [100] slabs of thickness 8.8 and 12.7~mm, using specularly reflecting faces represented by image sources. The incident plane pulse has a rectangular temporal profile of duration $t_p=0.3~\mu$s and is spatially smoothed over 50~$\mu$m. The arrival speed is defined as $d/(t_{\rm pk}-t_p/2)$, where $d$ is the propagation distance and $t_{\rm pk}$ is the time of the first temperature maximum exceeding 30\% of the largest peak and corresponding to a speed above 1.3~km/s. Calculations use the $17^3$ phonon mesh.

At the same nominal temperatures, the calculated peak arrival speeds are 7--10\% below the measured values at 15--18~K [Fig.~\ref{fig:S_experiments}(b)]. The calculations nevertheless reproduce the characteristic propagation speeds and their temperature dependence without fitting parameters to the heat-pulse measurements. At 8.8~mm, the fast temperature arrival persists through 18~K, but at 19~K the remaining maximum corresponds to approximately 0.4~km/s. At 12.7~mm, the fast arrival disappears between 17 and 18~K. This loss occurs approximately 1~K below the highest temperature at which a second-sound peak was measured, although the numerical boundary depends on the criterion used to identify the fast arrival.

The light-scattering experiment of Pohl and Irniger probes a grating period of 295~$\mu$m~\cite{pohl1976observation}. We calculate the frequency-dependent temperature response $\tilde F_T(q,i\omega)$ over 0--10~MHz and determine the resonance speed $\omega_{\rm res}/q$ from the peak of $\operatorname{Re}\tilde F_T$. The peak of $|\tilde F_T|$ provides an alternative value corresponding to the optimal detection phase. We also identify the leading complex-conjugate temperature-pole pair $p$, whose phase speed $|\operatorname{Im}p|/q$ and damping rate $-\operatorname{Re}p$ are independent of detection phase.

At 17~K, the calculated resonance speed is 2.19~km/s, close to the measured 2.16~km/s at 17.0~K [Fig.~\ref{fig:S_experiments}(c)], without fitting any parameters to the light-scattering experiment. The measured resonance speed exceeds the G--K hydrodynamic speed $v_h$ by approximately 16\%. At higher temperatures, however, the agreement deteriorates: the measured speed decreases toward and below $v_h$, reaching 1.56~km/s at 21.5~K, whereas the calculated speed remains 15--25\% higher.

The observation of a thermal resonance at 295~$\mu$m exposes a qualitative limitation of the independently parameterized G--K reduction. At 17~K, the G--K coefficients obtained from the $q=0$ collision spectrum (Table~S1) predict a return to overdamped temperature dynamics below the short-period edge of the G--K oscillatory window, 0.76~mm, whereas both the experiment and the full microscopic calculation exhibit an oscillatory response at a substantially shorter period. 

The experimental resonance was analyzed using Eq.~(3) of Ref.~\cite{pohl1976observation}, equivalent to the G--K temperature-response form [main-text Eq.~(17)] with $v_h=c/\sqrt{3}$ and $\gamma_q=c/\lambda_R$, where $c=3.15$~km/s is the first-sound velocity and $\lambda_R$ is the fitted damping length. Read out as the peak of $\operatorname{Re}\tilde F_T$ or of $|\tilde F_T|$, this response peaks at a speed no greater than $v_h$; a phase-rotated readout can peak above it. With the independently calculated G--K coefficients of Table~\ref{tab:S_gk}, the temperature response at 295~$\mu$m is overdamped throughout 16--21~K. The predicted short-period boundary of the oscillatory window lies between 0.6 and 1.4~mm, and the window closes near 21~K. At 17~K, the boundary is 0.76~mm, shifting to 0.42~mm under the linewidth-refinement diagnostic discussed below. These periods are already outside the microscopic validity range of the G--K reduction, so its short-period prediction represents an extrapolation of the single-relaxational-channel model. At 17~K, the calculated G--K relaxation rate $\gamma_q=\Gamma_h+\mu q^2\approx200~\mu$s$^{-1}$ exceeds the experimental resonance linewidth of approximately 1.9~$\mu$s$^{-1}$ in angular-frequency units by about two orders of magnitude.

\emph{Resonance damping and dispersion.}
Although the calculated resonance frequency agrees closely with experiment near 17~K, its linewidth is substantially larger. The leading temperature-pole damping increases from 5.3 to 10.9~$\mu$s$^{-1}$ over 15--21~K, whereas a single-pole fit to the measured 17.3-K resonance gives 0.74~$\mu$s$^{-1}$ [inset of Fig.~\ref{fig:S_experiments}(c)]. The calculated full width at half maximum is 2.1--2.2~MHz at 15--17~K, compared with the measured 0.30~MHz. This discrepancy is not attributable to the resonance-fitting procedure or detection phase: applying the same single-pole fit to the calculated response recovers its leading-pole damping within 2\%, and varying the detection phase does not produce a comparably narrow calculated resonance.

The calculation also predicts stronger spatial dispersion than suggested by the available measurements. At 17~K, the calculated resonance phase speed increases from 1.75~km/s at a 10-mm period to approximately 2.2~km/s below 0.5~mm [Fig.~\ref{fig:S_experiments}(d)]. By contrast, the measured heat-pulse and light-scattering speeds, which probe substantially different length scales, are similar: 1.79 and 1.82~km/s at 19~K. Because the heat-pulse calculations account for the experimental propagation geometry, this comparison suggests that the calculated dispersion between millimeter and submillimeter scales is stronger than observed.

To examine sensitivity to phonon linewidths, we perform a diagnostic refinement in which a denser integration mesh approximately doubles selected phonon linewidths used to update the even-parity sector of the collision operator. Incorporating these refined linewidths into the even sector of the $17^3$ kinetic generator reduces the G--K nonlocal coefficient $\mu$ from 0.444 to 0.245~m$^2$/s. However, it decreases the calculated resonance speed at 295~$\mu$m by only 3\% and does not reduce the fitted damping rate. The principal effect on the G--K oscillatory window is to shift its short-period boundary at 17~K from 0.76 to 0.42~mm, while the long-period boundary remains near 49~mm. This diagnostic does not replace the collision operator used in the main calculations.

\section{First-principles results for NaF}
\label{sec:S_reorg}

\subsection{Limits of the hydrodynamic reduction}
\label{sec:S_hydro}

\emph{G--K parameters.} Table~\ref{tab:S_gk} lists the G--K parameters of main-text Appendix~C\,2, taken from the odd-parity mode with the largest momentum overlap. These $q=0$ values define every G--K reference in this work. On heating, the G--K complex-pole window closes at 21.0~K. Up to 20~K the hydrodynamic mode carries at least 97\% of $\kappa$. Changing the momentum-overlap threshold from $H_h=0.6$ to $0.4$ shifts $T^*$ from 22.6 to 23.4~K.

\begin{table}[htbp]
\caption{G--K parameters of NaF from the $q=0$ collision spectrum (main-text Appendix~C\,2): bulk diffusivity $D_0$, hydrodynamic rate $\Gamma_h$, nonlocal coefficient $\mu$ and $v_h=(D_0\Gamma_h)^{1/2}$, for the odd-parity mode with the largest momentum overlap. $^\dagger$Above $T^*=22.9$~K no mode carries most of the momentum, and the values are formal continuations of the reduction.}
\label{tab:S_gk}
\begin{ruledtabular}
\begin{tabular}{lcccc}
$T$ (K) & $D_0$ (m$^2$/s) & $\Gamma_h$ ($\mu$s$^{-1}$) & $\mu$ (m$^2$/s) & $v_h$ (km/s) \\
\hline
16 & 17.8 & 0.196 & 0.583 & 1.87 \\
17 & 7.29 & 0.472 & 0.444 & 1.86 \\
18 & 3.30 & 1.03 & 0.364 & 1.84 \\
19 & 1.63 & 2.03 & 0.329 & 1.82 \\
20 & 0.871 & 3.72 & 0.347 & 1.80 \\
21 & 0.495 & 6.31 & 0.473 & 1.77 \\
22 & 0.297 & 9.88 & 1.00 & 1.71 \\
23$^\dagger$ & 0.187 & 14.0 & 3.25 & 1.62 \\
25$^\dagger$ & 0.0833 & 35.8 & 1.34 & 1.73 \\
30$^\dagger$ & 0.0178 & 151 & 0.352 & 1.64 \\
\end{tabular}
\end{ruledtabular}
\end{table}

\emph{Onset of the streaming pair.}\label{sec:S_ep} At 17~K, above $L_{\rm pair}$, the hydrodynamic rate follows $\Gamma_h+\mu q^2$ within 6\% with no fitted parameter. The two leading constitutive poles merge at $L_{\rm pair}=2.95$~mm on the $17^3$ mesh [main-text Fig.~2(d)]. Close to the merger their individual residues reach $1.01\pm3.37i$ in units of $\sum_nR_n$. A merger occurs on every mesh from $11^3$ to $21^3$ at every temperature from 15 to 22~K. At 17~K, $L_{\rm pair}$ decreases from 3.84~mm on $11^3$ to 2.88~mm on $21^3$, by 0.7\% in the last step. The two-mode estimate of main-text Appendix~A\,2, evaluated with the even mode $\iota$ most strongly coupled to $h$ (on the $17^3$ mesh $\varepsilon_\iota=4.0~\mu$s$^{-1}$ and $|\hat{\mathbf q}\cdot\boldsymbol\Pi_{h\iota}|=10.2$~\AA/ps), gives 3.6, 3.0 and 3.0~mm on the $17^3$, $19^3$ and $21^3$ meshes, against computed values of 2.95, 2.90 and 2.88~mm. The branch itself merges with the slowest even mode, so the agreement reflects the scale of the rate gap and coupling rather than the identity of a single partner mode. The refined phonon linewidths of Sec.~\ref{sec:S_second_sound} shorten $L_{\rm pair}$ to 2.29~mm, showing the sensitivity of its position to these rates. A merger remains present for every tested change of the rates.

\begin{figure}[htbp]
\centering
\includegraphics[width=0.4\textwidth]{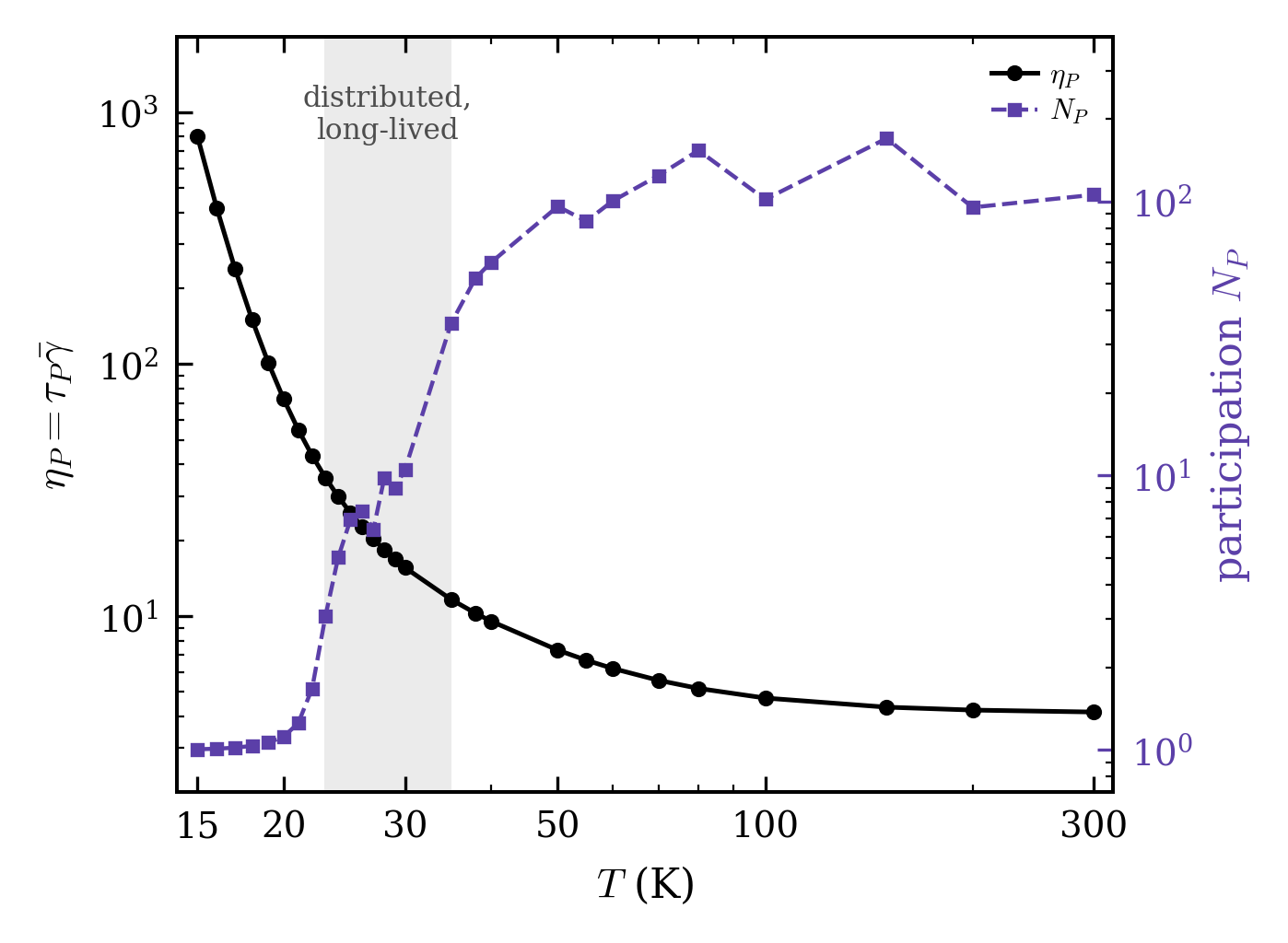}
\caption{Momentum persistence versus temperature at $q=0$, from the collision eigenmodes: momentum lifetime in mean collision times $\eta_P$ (left axis) and participation $N_P$ (right axis) [Eqs.~\eqref{eq:S_wv_CP} and~\eqref{eq:S_wv_tauP}]. Shading marks distributed but long-lived momentum, 22.9--35~K.}
\label{fig:S_hydro}
\end{figure}

\emph{Momentum correlation.} The momentum autocorrelation measures how quickly momentum correlations decay,
\begin{equation}
\begin{aligned}
&C_P(q,t)=\mathbf d_x^{\mathsf T}e^{-\boldsymbol\Xi(q)t}\mathbf d_x=\sum_nc_ne^{-a_nt},\\
&c_n=\langle d_x|\psi_n^{\mathrm R}\rangle\langle\psi_n^{\mathrm L}|d_x\rangle,\qquad\sum_nc_n=1,
\end{aligned}
\label{eq:S_wv_CP}
\end{equation}
where $\mathbf d_x$ represents the normalized crystal-momentum vector $|d_x\rangle$ of main-text Appendix~C\,2 in the working kinetic basis, $a_n$ runs over all eigenvalues of $\boldsymbol\Xi$ (the constitutive poles and the dark modes), $|\psi_n^{\mathrm R}\rangle$ and $\langle\psi_n^{\mathrm L}|$ are its right and left eigenvectors (main-text Sec.~II\,A), and $c_n$ are the biorthogonal residues (pairs summed). At $q=0$, $c_n=H_n$, so the largest residue is $H_h$, and above $T^*$ no single mode carries most of the momentum. From $C_P$ we define the participation $N_P=(\sum|c|)^2/\sum|c|^2$, the momentum lifetime
\begin{equation}
\tau_P=\int_0^\infty C_P\,dt=\mathbf d_x^{\mathsf T}\boldsymbol\Xi^{-1}\mathbf d_x,
\label{eq:S_wv_tauP}
\end{equation}
and its value in units of the mean collision time, $\eta_P=\tau_P\bar\gamma$ with $\bar\gamma$ the heat-capacity mean RTA rate of main-text Appendix~C. Figure~\ref{fig:S_hydro} gives the values at $q=0$, where they follow directly from the collision eigenmodes, $C_P(t)=\sum_\lambda H_\lambda e^{-\gamma_\lambda t}$ and $\tau_P=\sum_\lambda H_\lambda/\gamma_\lambda$. Between about 23 and 35~K the momentum is distributed among several modes, and its correlation persists for more than ten mean collision times; above 35~K it is spread over tens to more than a hundred modes, and at 100--300~K $\eta_P$ is about 4. At finite $q$, $C_P$ also decays through streaming within the energy-orthogonal space; the $q=0$ values isolate the collision contribution.

\subsection{Redistribution of heat-current weight at 17~K}
\label{sec:S_17K_evolution}

\label{sec:S_17K} 
\begin{figure}[htbp]
\centering
\includegraphics[width=0.8\textwidth]{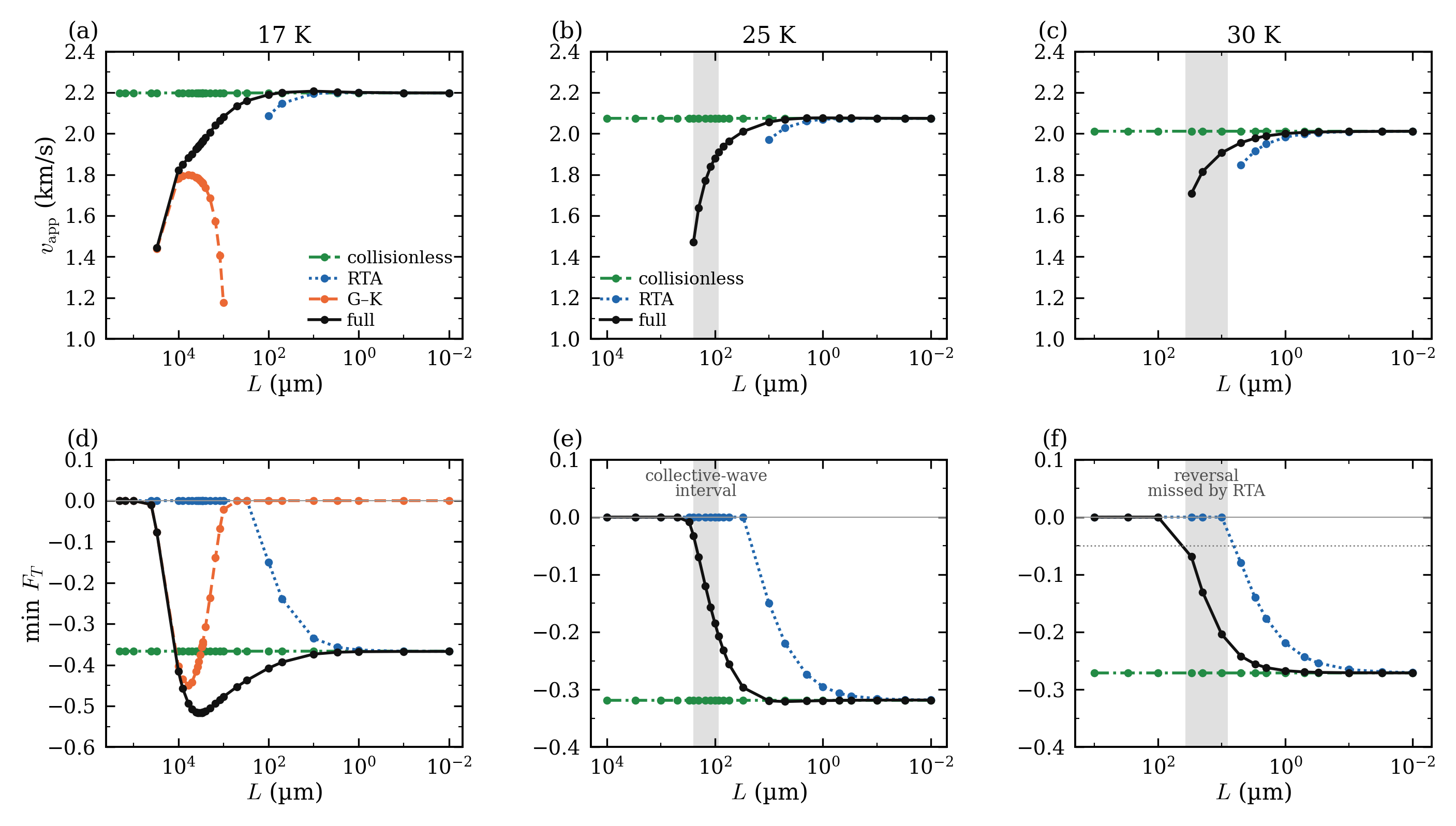}
\caption{Wave-vector dependence of transient-grating observables at 17, 25 and 30~K (columns), with $L$ decreasing to the right. (a)--(c) Apparent speed $v_{\rm app}=L/(2t_{\min})$, where $t_{\min}$ is the first minimum of $F_T$, shown when that minimum is below $-0.02$. (d)--(f) Minimum $F_T$ over $0\le tqv_{\rm rms}/(2\pi)\le4$, with responses that never become negative plotted at zero. G--K is shown at 17~K only. Shading in (b,e) marks the collective-wave interval. Shading in (c,f) marks sampled periods where the full response falls below $-0.05$ (dotted line) and RTA does not.}
\label{fig:S_qseries}
\end{figure}

The 17-K grating-period scan [Fig.~\ref{fig:S_qseries}(a,d)] extends main-text Fig.~3(e)--(h). Among the scan periods, the full response reverses sign at 44.9~mm and remains nonnegative at 50.3~mm; bisection places the onset at 48~mm (Table~\ref{tab:S_bounds}). The full and G--K responses show opposite apparent-speed trends with decreasing period [Fig.~\ref{fig:S_qseries}(a)]. Table~\ref{tab:S_bounds} lists the streaming crossover and RTA validity boundary marked in main-text Fig.~3(i). Below the short-period edge of the G--K window the full response remains strongly oscillatory, and from about $100~\mu$m down its apparent speed and minimum approach the collisionless reference.

\subsection{Transport without an isolated momentum sector}
\label{sec:S_collective}

\emph{Collective-wave screening.}\label{sec:S_screen} The collective thermal wave at 25~K is identified by testing whether the leading temperature-pole pair is underdamped and dominates a narrow frequency distribution. The three diagnostics are the pair weight $\pi_{\rm pair}=(|r_+|+|r_-|)/\sum_k|r_k|$ of the largest conjugate pair, normalized over individual poles; the damping ratio $|\operatorname{Im}p|/|\operatorname{Re}p|$ of that pair; and the relative frequency spread, the $|r_k|$-weighted relative standard deviation of $|\operatorname{Im}p_k|$ over underdamped poles ($|\operatorname{Im}p_k|>|\operatorname{Re}p_k|$). The pair passes the screening when $\pi_{\rm pair}\ge0.3$, $|\operatorname{Im}p|\ge|\operatorname{Re}p|$ and the spread is at most 0.2. A collective thermal wave passes these three tests and a fourth, continuous dispersion of the tracked pair across neighboring periods. The map share $P_{\rm pair}$ (main-text Sec.~II\,D) follows a different convention: $\pi_{\rm pair}$ adds the residue magnitudes of the two members over individual poles, whereas $P_{\rm pair}$ takes the magnitude of the summed pair residue over grouped channels, and the two shares have no general ordering. The second-sound class in the regime map requires a majority temperature-pole pair and sign reversal, whereas the thermal-wave screening identifies a distinguishable underdamped pair without requiring majority weight. Consequently, most of the 25-K thermal-wave interval is classified as a damped oscillatory response in Fig.~6(b).

\begin{figure}[htbp]
\centering
\includegraphics[width=0.8\textwidth]{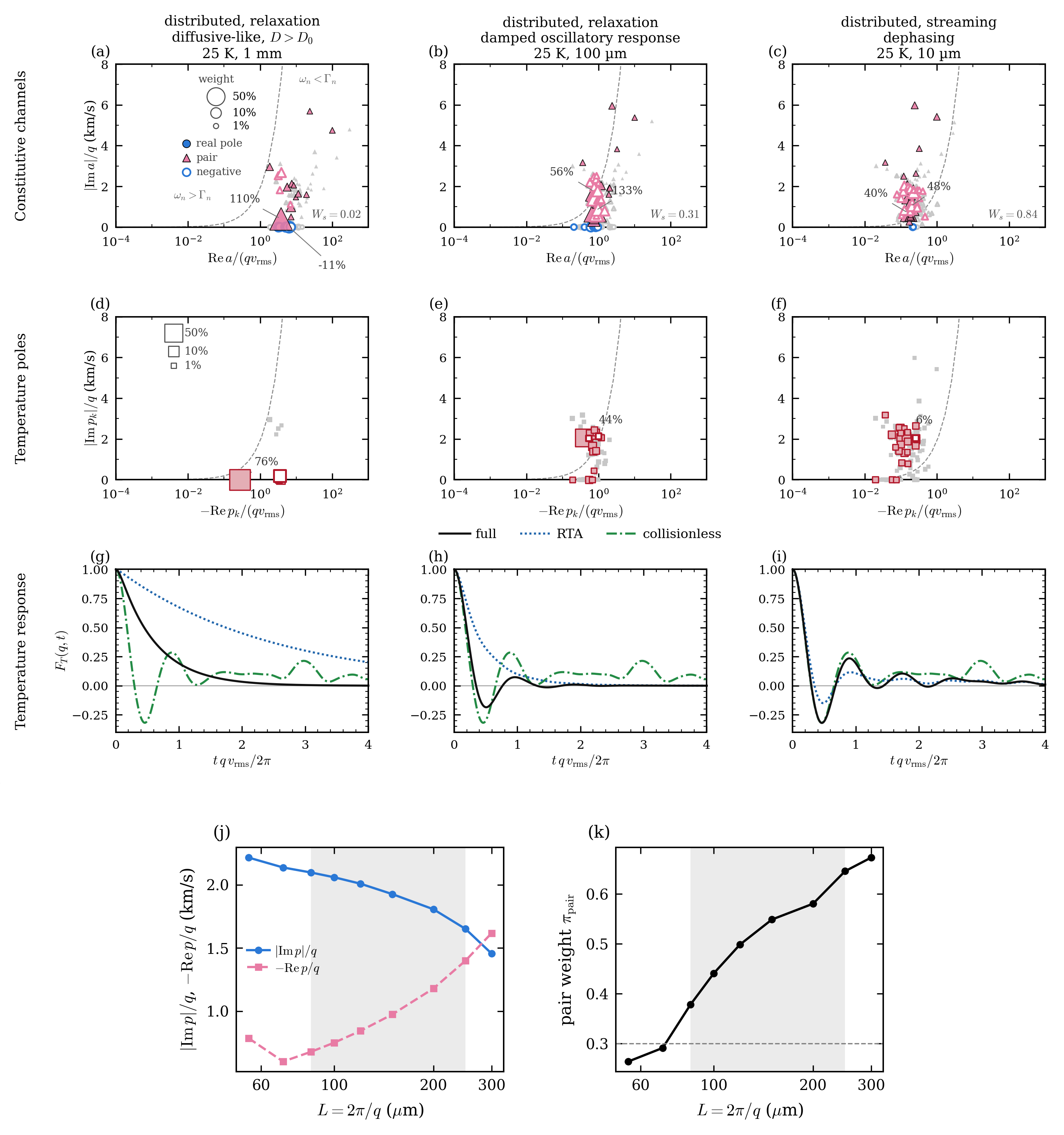}
\caption{Collective thermal wave at 25~K. (a)--(i) Three periods, $L=1$~mm, $100~\mu$m and $10~\mu$m. (a)--(c) Constitutive channels, with the marker conventions of main-text Fig.~4(a)--(d), and the values of $W_s$. (d)--(f) Temperature poles, sized by their share of $\sum_k|r_k|$ with conjugate partners combined; open squares indicate a residue with negative real part, shares below 0.2\% are omitted, and dashed curves mark $|\operatorname{Im}p_k|=|\operatorname{Re}p_k|$. (g)--(i) Normalized temperature correlations from the full theory, RTA, and collisionless limit. (j,k) Period scan of the leading temperature-pole pair: (j) phase speed $|\operatorname{Im}p|/q$ and damping scale $-\operatorname{Re}p/q$; (k) pair weight $\pi_{\rm pair}$, with the 30\% threshold. Shading in (j,k): the sampled interval 85--$250~\mu$m, which passes all four collective-wave tests of Sec.~\ref{sec:S_screen}.}
\label{fig:S_wave25}
\end{figure}

\emph{25 K: collective thermal wave.}\label{sec:S_dispersion} The leading temperature-pole pair at 25~K, which carries 44\% of $\sum_k|r_k|$ at $100~\mu$m [Fig.~\ref{fig:S_wave25}(b,e,h)], is tracked by its eigenvector in the fixed $q=0$ basis, with overlaps of 0.94--0.995 between adjacent lengths from 70 to $250~\mu$m [Fig.~\ref{fig:S_wave25}(j,k)]. Over 85--$250~\mu$m it carries at least 38\% of $\sum_k|r_k|$, is underdamped, and its relative frequency spread stays below 0.17. The interval ends where the pair weight falls below 0.3 at $70~\mu$m and where the pair becomes overdamped at $300~\mu$m. In scaled time, the full response depends on the period although the collisionless reference does not [Fig.~\ref{fig:S_qseries}(b,e)]. The persistence of this underdamped temperature-pole pair without an isolated momentum-organized sector demonstrates that oscillatory temperature dynamics need not originate from conventional phonon hydrodynamics.

\emph{30 K: distributed streaming.}\label{sec:S_30K} At $10~\mu$m the largest temperature-pole pair carries about 6\% of $\sum_k|r_k|$, and unlike at 25~K no period shows a distinguishable temperature-pole pair. The full response falls below $-0.05$ at periods of $30~\mu$m and shorter, whereas RTA stays positive over 10--$30~\mu$m [Fig.~\ref{fig:S_qseries}(f)] and reverses sign at shorter periods [main-text Fig.~4(k)]. At 100~nm the full correlation is close to the collisionless one, but the period lies above $L_{\rm cl}$, so the map labels it dephasing.

\emph{300 K: streaming without temperature sign reversal.}\label{sec:S_300K} At 300~K the first minimum of the collisionless correlation is $+0.13$ (against $-0.27$ at 30~K): 58\% of the heat capacity lies in phonons with $|v_{\alpha L}|<1$~km/s, against 30\% at 30~K. The first minimum of the collisionless correlation crosses zero at 69.8~K on the $17^3$ mesh and at 72.0~K on a $61^3$ mesh, and its depth changes by less than 0.01 between them. Later in the window, near $tqv_{\rm rms}/2\pi=2.5$, the 300-K correlation returns to within $\pm0.01$ of zero, with a sign that depends on the mesh.

\section{Transport crossovers and regime maps}
\label{sec:S_crossmap}

\subsection{Transport boundaries and analytical estimates}
\label{sec:S_map}
\label{sec:S_windows}

\begin{figure}[htbp]
\centering
\includegraphics[width=0.8\textwidth]{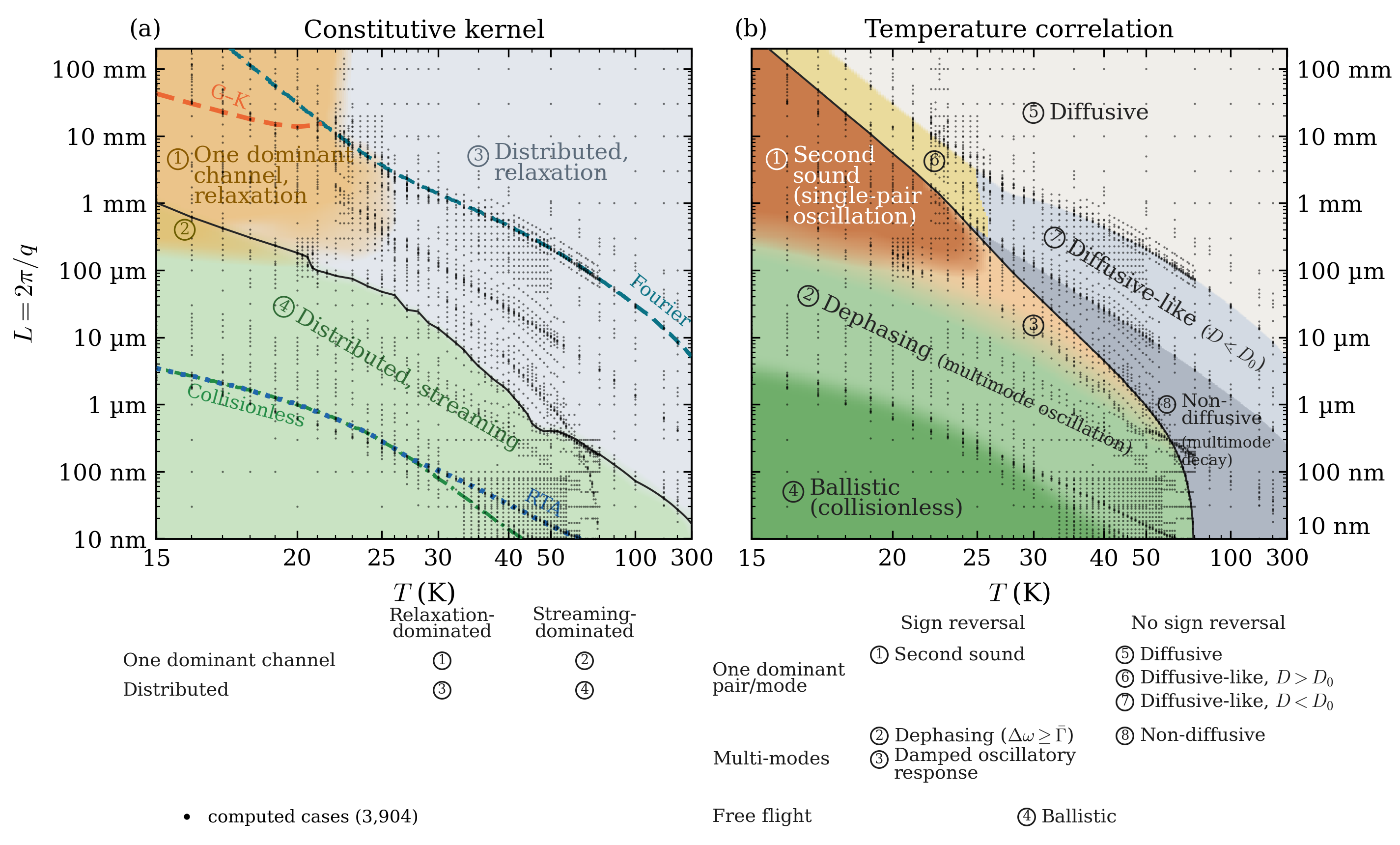}
\caption{Sampling of the maps of main-text Fig.~6, drawn as in the main text. Dots: the 3,904 computed cases, including the calculations that locate the lines and class edges.}
\label{fig:S_map_cases}
\end{figure}

The maps of main-text Fig.~6 sample 15--300~K and 10~nm--1~m on a nonuniform grid in $T$ and $\log L$, with additional temperatures near the crossovers. Stored constitutive and temperature spectra determine the channel and response classes, and additional calculations locate each line by bisection or local scans (Fig.~\ref{fig:S_map_cases}). The constitutive validity lines are the bisected periods at which each reduced kernel departs from the full kernel by 1\%; where an edge lies beyond the computed periods, as for $L_F$ at 15 and 17~K, the expression of main-text Appendix~C is used. The sign-reversal, $P_D=1/2$ and diffusive edges are log-midpoints of bisected brackets with ratios of at most 1.03. Mesh checks cover $L_{\rm pair}$ at 17~K (Sec.~\ref{sec:S_ep}), the 25- and 30-K classes (Sec.~\ref{sec:S_mesh}) and the collisionless sign-reversal crossing (Sec.~\ref{sec:S_300K}); the other edges are computed on the $17^3$ mesh, and their sensitivity to scattering rates
and force constants has not been quantified.

\emph{Comparison measures.} Each reduced description $X$ is compared with the full theory with relative tolerance $\delta=1\%$. Fourier and G--K hold at long periods, where the temperature response samples the kernel over the frequency band relevant to temperature relaxation. They are measured by the band-limited kernel error
\begin{equation}
e_X=\max_{0\le\omega\le\omega_T}\left|\frac{\tilde Z_L(q,i\omega)}{\tilde Z_L^X(q,i\omega)}-1\right|,
\label{eq:S_band_error}
\end{equation}
where $\omega_T$ is the median frequency of the temperature power spectrum $|\tilde F_T(q,i\omega)|^2$. This band also excludes the instantaneous weight of the faster channels, which G--K omits by construction. RTA and collisionless streaming hold at short periods, where the response has no diffusive band. They are measured in the scaled time $\theta=qv_{\rm rms}t$ over the first oscillation of the collisionless temperature correlation, $0\le\theta\le\theta_w$, where $\theta_w$ is the first maximum of $F_T^{\rm cl}$ after its first minimum. For a curve $x$ and a reference $x_{\rm ref}$,
\begin{equation}
d=\frac{\|x-x_{\rm ref}\|_w}{\|x_{\rm ref}\|_w}=\left[\frac{\int_0^{\theta_w}(x-x_{\rm ref})^2\,d\theta}{\int_0^{\theta_w}x_{\rm ref}^2\,d\theta}\right]^{1/2},
\label{eq:S_window_distance}
\end{equation}
with $x=z\equiv Z_L(q,t)/Z_L(q,0)$ for the kernel distances $d_{Z,w}^X$ that define $L_{\rm RTA}$ and $L_{\rm cl}$, and $x=F_T$ for the temperature distance $d_{T,w}^{\rm cl}$ of the ballistic check (Sec.~\ref{sec:S_variables}). Reference kernels follow from the reference temperature correlations through energy conservation, $\partial_\theta F_T=-\int_0^\theta z(\theta-\theta')F_T(\theta')\,d\theta'$, solved as a Volterra equation of the second kind.

\emph{Accuracy of the boundary expressions.} The expressions of main-text Appendix~C reproduce the computed 1\% edges as follows. $L_F$ agrees within 5\% up to 40~K; at higher temperature the neglected $q^4$ terms make it longer, by 20\% at 300~K. $L_{\rm GK}$ agrees within 3\% at 15--22~K. The first-order estimate overestimates $L_{\rm RTA}$ by 9--12\% below 30~K; with $z_2^{\rm od}$ included it agrees within 1.7\% at 15--60~K. For collisionless streaming, the net decay rate of the current is $\bar\gamma_J^{\rm full}\approx0.18\,\bar\gamma_J$ at 15--20~K and $0.5\,\bar\gamma_J$ at 40~K; the expansion around free streaming does not converge at the 1\% edge below about 30~K (main-text Appendix~C\,4), and the RTA-resummed expression agrees with the computed $L_{\rm cl}$ within 1.2\% at 15--46~K.

\emph{Boundary values.} Table~\ref{tab:S_bounds} lists the lines of main-text Fig.~6 at selected temperatures. In panel (a), the largest channel share approaches one-half at millimeter periods near 25~K, and the class with one dominant streaming channel is confined to 15--19~K. The Fourier boundary $L_F$ is set by the memory length $\ell_m$ below about 25~K and by the static length $\ell_s$ above, and the G--K boundary meets it at 21.3~K. The collisionless boundary satisfies $2\pi\delta v_{\rm rms}/\bar\gamma_J<L_{\rm cl}<2\pi\delta v_{\rm rms}/\bar\gamma_J^{\rm full}$ at every computed temperature. The relation between these validity domains and the kernel classes is discussed in main-text Sec.~IV\,B.

\begin{table*}[htbp]
\caption{Lines of main-text Fig.~6 at selected temperatures, rounded to two significant figures. Panel (a): streaming dominates below $L_s$; Fourier and G--K hold above $L_F$ and $L_{\rm GK}$, and RTA and collisionless streaming below $L_{\rm RTA}$ and $L_{\rm cl}$ (1\% tolerance). Panel (b): the temperature correlation reverses sign below the listed period; without sign reversal, the slowest temperature channel carries at least half the amplitude above the edge $P_D=1/2$, and the response is diffusive above the diffusive edge. $^a$Expression of main-text Appendix~C; the edge lies beyond the computed periods. G--K: no controlled reduction above $T^*$ (\textit{n.c.}). \textit{none}: the line does not occur at this temperature. $<10$~nm, $>1$~m: outside the mapped range.}
\label{tab:S_bounds}
\begin{ruledtabular}
\begin{tabular}{rcccccccc}
 & \multicolumn{5}{c}{Panel (a)} & \multicolumn{3}{c}{Panel (b)} \\
$T$ (K) & $L_s$ & $L_F$ & $L_{\rm GK}$ & $L_{\rm RTA}$ & $L_{\rm cl}$ & Sign reversal & $P_D=1/2$ & Diffusive \\
\hline
15 & 1.0~mm & 1.6~m$^a$ & 43~mm & 2.6~$\mu$m & 3.4~$\mu$m & 330~mm & \textit{none} & $>1$~m \\
17 & 0.42~mm & 0.25~m$^a$ & 23~mm & 1.4~$\mu$m & 2.1~$\mu$m & 48~mm & \textit{none} & 0.25~m \\
20 & 0.18~mm & 30~mm & 14~mm & 0.61~$\mu$m & 1.0~$\mu$m & 5.5~mm & \textit{none} & 30~mm \\
23 & 75~$\mu$m & 7.1~mm & \textit{n.c.} & 0.32~$\mu$m & 0.48~$\mu$m & 1.0~mm & \textit{none} & 6.8~mm \\
25 & 48~$\mu$m & 3.7~mm & \textit{n.c.} & 0.23~$\mu$m & 0.29~$\mu$m & 0.34~mm & 0.41~mm & 3.1~mm \\
30 & 14~$\mu$m & 1.4~mm & \textit{n.c.} & 0.10~$\mu$m & 80~nm & 47~$\mu$m & 0.12~mm & 1.1~mm \\
40 & 1.6~$\mu$m & 0.46~mm & \textit{n.c.} & 32~nm & 14~nm & 4.5~$\mu$m & 26~$\mu$m & 0.44~mm \\
50 & 0.41~$\mu$m & 0.21~mm & \textit{n.c.} & 16~nm & $<10$~nm & 0.95~$\mu$m & 10~$\mu$m & 0.20~mm \\
100 & 72~nm & 30~$\mu$m & \textit{n.c.} & $<10$~nm & $<10$~nm & \textit{none} & 1.4~$\mu$m & 30~$\mu$m \\
300 & 17~nm & 5.1~$\mu$m & \textit{n.c.} & $<10$~nm & $<10$~nm & \textit{none} & 0.25~$\mu$m & 5.0~$\mu$m \\
\end{tabular}
\end{ruledtabular}
\end{table*}

The constitutive regions and temperature-response classes do not have a one-to-one correspondence. In panel (b), the majority-pair second-sound class ends at 25~K and dephasing at 69~K. The sign-reversal region closes at 69.8~K, where the collisionless minimum crosses zero; above 56~K its short-period edge falls from about $0.4~\mu$m to 18~nm at 69~K.

\subsection{Spectral classification and validation}
\label{sec:S_variables}
\label{sec:S_decay}

Map shares are normalized by the sum of the grouped channel weights (main-text Sec.~II\,D). The streaming index and the dominance share use the coalescence grouping: poles whose positions differ by less than 2\% and whose residues cancel strongly, $|\sum R_n|\le0.2\sum|R_n|$, form one channel, so every channel still has a defined $\Gamma_n$ and $\omega_n$. Near a coalescence, individual residues grow with opposite signs ($\sum_n|R_n|/\sum_nR_n$ up to about 7), and without the grouping $\hat w_1$ and $W_s$ show spurious dips or spikes at the coalescence. At 15--22.5~K the stored temperature poles are sparse at long periods, and the second-sound region is continued up to the reversal line; every computed case between its lower edge and the reversal line (160 cases) has a dominant pair and sign reversal.

The ballistic class is the collisionless domain $L<L_{\rm cl}$ of panel (a) and, in the mapped range ($L\ge10$~nm), extends to about 43~K. 

At the two sampled periods bracketing $L_{\rm cl}$, the temperature distances $d_{T,w}^{\rm cl}$ are 0.23--0.26\% at 17~K ($L=1.99$--$2.28~\mu$m), 0.33--0.37\% at 20~K ($0.949$--$1.10~\mu$m), 0.76--0.85\% at 30~K (74.0--86.0~nm), and 0.80--0.83\% at 40~K (13.3--14.0~nm). These are evaluations at nearby sampled periods, using the window and reference normalization of Sec.~\ref{sec:S_crossmap}.

\emph{Decay amplitude.} The decay amplitude $P_D$ is the signed residue of the slowest temperature channel, with $\sum_kr_k=1$, rather than its share of $\sum_kw_k$; where all residues are positive the two coincide. Near the sign-reversal line, faster temperature poles that nearly coalesce carry large residues of opposite sign, which cancel in $F_T$ but inflate $\sum_kw_k$, and they lie farther apart than the 2\% grouping distance. The signed residue is unaffected by this cancellation. The edge $P_D=1/2$ meets the reversal line between 24 and 24.5~K. The diffusive class uses the same slow pole. Up to 27~K its edge is set by the amplitude condition, and up to 24~K it lies within 10\% of $L_F$; from 28~K it is set by the rate condition. At 25.5--30~K the slow pole already decays within 1\% of the Fourier rate while faster poles with cancelling residues still shape the early-time response, so there the class extends to shorter periods than the 1\% agreement of $F_T$ with Fourier. At other temperatures the pole edge lies at equal or longer periods. The slow-pole approximation of main-text Appendix~C\,1 was tested on the 325 nonreversing cases at 21--49~K whose slowest temperature channel is a dominant real pole. It reproduces the exact decay rate within 1\% in 283 cases, against 145 for the static factor $K$ alone. The largest errors, up to 33\%, occur next to the sign-reversal line, where $\epsilon_{\rm mem}$ reaches 0.22--0.27 and the first-order expansion in the slow-pole rate is least accurate. At 25~K the memory and static-nonlocality lengths $\ell_m$ and $\ell_s$ are both about $50~\mu$m, so the two leading long-wavelength corrections nearly balance.

\bibliographystyle{apsrev4-2}
\bibliography{unified_theory}